\documentclass[12pt]{article}
\pdfoutput=1
\usepackage{amssymb, amsmath,amsfonts}
\usepackage{graphicx}
\usepackage[pdftex,bookmarks=true,colorlinks,linkcolor=blue,urlcolor=blue,citecolor=blue]{hyperref}

\makeatletter\@addtoreset{equation}{section}\makeatother

\def\be{\begin{equation}}
\def\ee{\end{equation}}
\def\bea{\begin{eqnarray}}
\def\eea{\end{eqnarray}}

\makeatletter\@addtoreset{equation}{section}\makeatother

\newcommand{\preprint}[1]{\begin{table}[t]  
             \begin{flushright}               
             {#1}                             
             \end{flushright}                 
             \end{table}}                     
\renewcommand{\title}[1]{\vbox{\center\LARGE{#1}}\vspace{5mm}}
\renewcommand{\author}[1]{\vbox{\center#1}\vspace{5mm}}
\newcommand{\address}[1]{\vbox{\center\em#1}}

\begin{document}

\unitlength = .8mm

\begin{titlepage}
\vspace{.5cm}
\preprint{IFT-UAM/CSIC-16-022}

\begin{center}
\hfill \\
\hfill \\
\vskip 1cm

\title{Negative magnetoresistivity in holography}

\vskip 0.5cm
 {Ya-Wen Sun$^{a}$}\footnote{Email: {\tt yawen.sun@csic.es}} and
 {Qing Yang$^{a, b}$}\footnote{Email: {\tt yangqing@itp.ac.cn}}

\address{${}^a$Instituto de F\'\i sica Te\'orica UAM/CSIC, C/ Nicol\'as Cabrera 13-15,\\
Universidad Aut\'onoma de Madrid, Cantoblanco, 28049 Madrid, Spain}
\address{${}^b$Institute of Theoretical Physics, Chinese Academy of Sciences,\\ Beijing 100190, China}

\end{center}

\vskip 1.5cm

\abstract{Negative magnetoresistivity is a special magnetotransport property associated with chiral anomaly in four dimensional chiral anomalous systems, which refers to the transport behavior that the DC longitudinal magnetoresistivity decreases with increasing magnetic field. We calculate the longitudinal magnetoconductivity in the presence of backreactions of the magnetic field to gravity in holographic zero charge and axial charge density systems with and without axial charge dissipation. In the absence of axial charge dissipation, we find that the quantum critical conductivity grows with increasing magnetic field when the backreaction strength is larger than a critical value, in contrast to the monotonically decreasing behavior of quantum critical conductivity in the probe limit. With axial charge dissipation, we find the negative magnetoresistivity behavior. The DC longitudinal magnetoconductivity scales as $B$ in the large magnetic field limit, which deviates from the exact $B^2$ scaling of the probe limit result. In both cases, the small frequency longitudinal magnetoconductivity still agrees with the formula obtained from the hydrodynamic linear response theory, even in the large magnetic field limit.}

\vfill

\end{titlepage}

{\hypersetup{linkcolor=black}
\tableofcontents}


\section{Introduction}

In four dimensions, there exist systems of chiral fermions which possess chiral anomaly, including quark-gluon plasma, Dirac or Weyl (semi-) metals, see \cite{{Kharzeev:2015znc},{qi_review}} for recent reviews and references therein. In these systems, due to the existence of the chiral anomaly, there are several associated anomalous transport behaviors, including negative magnetoresistivity \cite{Nielsen:1983rb}, anomalous Hall effect, chiral magnetic effect \cite{Fukushima:2008xe}, chiral vortical effect \cite{Erdmenger:2008rm,{Banerjee:2008th}}, etc. Chiral magnetic and vortical effects have been studied extensively in holographic chiral anomalous systems \cite{{Erdmenger:2008rm},{Banerjee:2008th},{Son:2009tf},{Yee:2009vw},{Neiman:2010zi},{Landsteiner:2011cp},{Landsteiner:2011iq},{Jensen:2012kj},{Hou:2012xg}} via the gauge/gravity duality (see \cite{{Ammon:2015wua},{Nastase:2015wjb},{Zaanen}} for recent reviews). Meanwhile, anomalous Hall effect was proposed as an order parameter in the realization of a holographic quantum phase transition between a topological and a trivial semi-metal state \cite{Landsteiner:2015lsa,{Landsteiner:2015pdh}}.

Negative magnetoresistivity refers to the anomalous transport behavior of the longitudinal DC magnetoresistivity decreasing with increasing magnetic field, or the longitudinal DC magnetoconductivity increasing with increasing magnetic field in the presence of chiral anomaly, in contrast to the positive magnetoresistivity behavior for normal metal \cite{wannier}. Negative magnetoresistivity has been observed in several experiments during the last several years, including \cite{kimprl,{Li:2014bha},{Huang},{Zhang},{Xiong},{Yan},{Ong}}.

Using a linear response theory in the hydrodynamic regime \cite{linearresponse,{nernsteffect}}, it was shown in \cite{Landsteiner:2014vua} (and later generalized to Lifshitz spacetime in \cite{Roychowdhury:2015jha}) that the longitudinal DC magnetoconductivity in the presence of chiral anomaly is divergent, even in the zero density limit. Energy, momentum and axial charge dissipations are all needed to make it finite. At the zero charge and axial charge density limit, only axia charge dissipation is needed to have a finite DC longitudinal magnetoconductivity. At weak coupling, from the kinetic theory \cite{{Nielsen:1983rb},{sonspivak},{gorbar},{Goswami:2015uxa}}, it was calculated that at small $B\ll T^2$, the DC longitudinal magnetoconductivity has a $B^2$ behavior while at large $ B \gg T^2$, it goes linearly in $B$. Negative magnetoresistivity behavior was also found in strongly coupled holographic chiral anomalous systems \cite{{Lifschytz:2009si},{Landsteiner:2014vua},{Jimenez-Alba:2014iia},
{Jimenez-Alba:2015awa}}. In \cite{Landsteiner:2014vua} we found that when there is no axial charge dissipation, the longitudinal magnetoconductivity indeed has a pole at $\omega=0$ which leads to a $\delta$-function in the real part of the longitudinal magnetoconductivity. The real part of the conductivity at zero frequency excluding the $\delta$-function is a monotonically decreasing function of $B$ and decreases from $\pi T$ at $B=0$ to $0$ at $B=\infty$. The coefficient in front of $i/\omega$ of the imaginary part is a monotonically increasing function of $B$ and increases from $B^2$ at $B=0$ to linear in $B$ behavior at large $B$. This kind of scaling behavior in a strongly coupled holographic system coincides with the weakly coupled result, and was also found in experiments \cite{kimprl}.

There are at least two ways to introduce axial charge dissipations into the holographic system \cite{Jimenez-Alba:2015awa}. The first is by explicitly breaking the $U(1)_A$ symmetry with a $U(1)_A$ charged scalar in the bulk which has a nonzero source at the boundary. The second way is to make the $U(1)_A$ gauge field massive \cite{{Gursoy:2014ela},{Jimenez-Alba:2014iia},{Iatrakis:2015fma}} so that there is no $U(1)_A$ gauge symmetry in the bulk anymore. The two ways are in fact equivalent in the following sense: the equations for the perturbations are the same for the two mechanisms at zero charge and axial charge density after choosing suitable gauges and substituting the mass of the $U(1)_A$ gauge field by the background scalar field. In fact the massive $U(1)_A$ case is one special limit of the explicit breaking case where the mass of the scalar field is chosen to be zero. With explicit breaking of the axial charge conservation symmetry, we found that the DC conductivity is composed of two terms and the non-constant term has an exact $B^2$ dependence on the magnetic field $B$. This qualitatively agrees with the experimental result of \cite{Li:2014bha}. In both of the holographic zero density systems with and without axial charge dissipation, the hydrodynamic results agree with the holographic results as long as $\tau_5$ is large enough to stay in the hydrodynamic regime while $B$ can be very large which is outside of the hydrodynamic regime.

Previous study of negative magnetoresistivity in holographic chiral anomalous systems focused on the probe limit, where the magnetic field cannot be very large so that backreactions of the magnetic field to gravity are not important. To study the large $B$ behavior more accurately, we need to take into account the backreaction effects of the magnetic field. In this paper, we study the holographic zero charge and axial charge density systems with the backreactions of the magnetic field.  This is also a first step towards the study of magnetotransport behavior in holographic finite charge and axial charge density chiral anomalous systems, where backreactions of guage fields should always be considered. In this paper we consider both the cases without and with axial charge dissipation for the zero density system. We find that in the case without axial charge dissipation, the small frequency longitudinal magnetoconductivity deviates from the probe limit at larger $B/T^2$ region. At $B/T^2\to \infty$, the imaginary part of the longitudinal magnetoconductivity coincides with the probe limit result while the real part diverges for backreaction strength larger than a critical value, in contrast to being zero in the probe limit. In the case with axial charge dissipation, at large $B/T^2$ the DC longitudinal magnetoconductivity becomes linear in $B$, which deviates from the exact $B^2$ behavior for the probe limit.

The rest of the paper is organized as follows. In section 2, we will calculate the longitudinal magnetoconductivity with backreactions to the gravity without axial charge dissipation at both finite and zero temperature. In section 3, we add axial charge dissipations and calculate the longitudinal magnetoconductivity at finite temperature. Section 4 is devoted to conclusion and discussions.
\section{Backreacted $U(1)_\text{V}\times U(1)_\text{A}$ holographic system with magnetic field: without axial charge dissipation}
\label{sec2}
In this section, we calculate the magnetoconductivity in the presence of backreactions of the magnetic field to the gravity at both zero charge density and zero axial charge density without introducing any dissipations and compare the result with the probe case, especially at the large magnetic field limit. We will consider the following action\footnote{We have set the curvature scale $L=1$.}
\bea
S=\int d^5x \sqrt{-g}\bigg[\frac{1}{2\kappa^2}\Big(R+12\Big)-\frac{1}{4 e^2}(\mathcal{F}^2+F^2)+\frac{\alpha}{3}\epsilon^{\mu\nu\rho\sigma\tau}A_\mu \Big(F_{\nu\rho} F_{\sigma\tau}+3 \mathcal{F}_{\nu\rho} \mathcal{F}_{\sigma\tau}\Big)\bigg]\,
\eea
with
\be
\mathcal{F}_{\mu\nu}=\partial_\mu V_\nu-\partial_\nu V_\mu\,,
~~F_{\mu\nu}=\partial_\mu A_\nu-\partial_\nu A_\mu\,
\ee
where the gauge fields $V_\mu$ and $A_\mu$ correspond to the vector and axial $U(1)$ currents respectively. $\kappa$ is the Newton constant, $e$ is the Maxwell coupling constant and $\alpha$ is the Chern-Simons coupling constant. Here we did not introduce any dissipation terms and according to the hydrodynamic formula in \cite{Landsteiner:2014vua} we will get a $\delta$-function at zero frequency which leads to an infinite DC magnetoconductivity.

The equations of motion are
\bea\label{eq:eomtwou1-1}
R_{\mu\nu}-\frac{1}{2}g_{\mu\nu}\Big(R-12 -\frac{\kappa^2}{2 e^2}(\mathcal{F}^2+F^2)\Big)
-\frac{\kappa^2}{e^2} \mathcal{F}_{\mu\rho}\mathcal{F}_{\nu}^{~\rho}-\frac{\kappa^2}{e^2} F_{\mu\rho}F_{\nu}^{~\rho}&=&0\\
\nabla_\nu \mathcal{F}^{\nu\mu}+2\alpha\epsilon^{\mu\tau\beta\rho\sigma} F_{\tau\beta}\mathcal{F}_{\rho\sigma}&=&0\,,\\
\label{eq:eomtwou1-2}\nabla_\nu F^{\nu\mu}+\alpha\epsilon^{\mu\tau\beta\rho\sigma} \big(F_{\tau\beta}F_{\rho\sigma}
+\mathcal{F}_{\tau\beta}\mathcal{F}_{\rho\sigma}\big)&=&0\,
\label{eq:eomtwou1-3}
\eea
We will first solve this system with a finite magnetic field and then consider perturbations on this background to get the longitudinal magnetoconductivity.

\subsection{Background solution at finite temperature}

We turn on a nonzero magnetic field at zero charge density and zero axial charge density. The ansatz for the background solutions are
\be
ds^2=- f(r) dt^2+ \frac{dr^2}{f(r)}+n(r)(dx^2+dy^2)+ h(r) dz^2,
\ee
and
\be
V_{\mu}=(0,~0,~B y,~0,~0),~~
A_{\mu}=(0,~0,~0,~0,~0).
\ee


The background equations of motion become
\bea
\frac{f''}{f}-\frac{n''}{n}+\Big(\frac{f'}{f}-\frac{n'}{n}\Big)\frac{h'}{2h}-\frac{\lambda B^2}{n^2f}&=&0\,\label{eqn001},\\
\frac{f''}{2f}+\frac{n''}{n}+\Big(\frac{f'}{f}-\frac{n'}{4n}\Big)\frac{n'}{n}+\Big(-6+\frac{\lambda B^2}{4n^2}\Big)\frac{1}{f}
&=&0\,\label{eqn002},\\
\Big(\frac{n'^2}{n^2}+\frac{2n'h'}{nh}\Big)f+\Big(\frac{h'}{h}+\frac{2n'}{n}\Big)f'+\frac{\lambda B^2}{n^2}-24&=&0\,\label{eqn003},
\eea
where $\lambda=\frac{2\kappa^2}{e^2}$ is a dimensionless constant, which represents the strength of backreactions. For this system, there are three second order equations and one first order equation coming from the Einstein's equations of motion and only three of them are independent. For convenience in numerics, we choose the three equations above to eliminate the second derivative of $h$ in the equations: (\ref{eqn001}) is a linear combination of the $tt$ and $xx$ components of the Einstein's equations of motion, (\ref{eqn002}) is the $zz$ component and (\ref{eqn003}) is the $rr$ component, which is a first order equation. Note that in the regime of classical gravity, $\kappa\ll 1$, while $\lambda$ can be arbitrarily large or small depending on the ratio of $\kappa/e$. At zero charge and axial charge density, these equations of motion coincide with the equations for the case with only one $U(1)$ gauge field \cite{D'Hoker:2009mm,{Janiszewski:2015ura}}.



One exact solution to this system is the BTZ$\times$ R$^2$ solution \be f=3 (r^2-1),~~n=\frac{B\sqrt{\lambda}}{2\sqrt{3}},~~h=r^2\ee with the $AdS_3$ radius being $1/\sqrt{3}$. However, we cannot find irrelevant deformations to flow this solution to asymptotic $AdS_5$ geometry and only marginal deformations can be found, which render the near horizon geometry no longer BTZ$\times$ R$^2$ any more. Thus at finite temperature, it is more convenient to directly expand the solutions at the horizon as follows
\be f\simeq 4\pi T(r-1)+\cdots, ~~~n\simeq n_0+\cdots,~~~h\simeq h_0+\cdots,\ee
where the near horizon parameter $T$ is the temperature of the background solution and the horizon radius $r_0$ has already been rescaled to $1$. The ``$\cdots$'' represents higher order expansions which can be determined order by order given the values of $T,~n_0$ and $h_0$.

The asymptotic $AdS_5$ boundary behaviors of the metric fields are
\be
f\simeq r^2\bigg{(}1+\frac{2 f_0}{r}+\frac{f_0^2}{r^2}-\frac{B^2\lambda \ln r}{6 r^4}-\frac{M}{r^4}+\cdots\bigg{)}
\ee
\be
n\simeq r^2\bigg{(}1+\frac{2 f_0}{r}+\frac{f_0^2}{r^2}+\frac{B^2\lambda \ln r}{12 r^4}+\frac{n_2}{r^4}+\cdots\bigg{)}
\ee
\be
h\simeq r^2 \bigg{(}1+\frac{2 f_0}{r}+\frac{f_0^2}{r^2}-\frac{B^2\lambda \ln r}{6 r^4}-\frac{2 n_2}{r^4}+\cdots\bigg{)},
\ee where $M$ and $n_2$ are parameters which are determined by the horizon data and $f_0$ can be eliminated by performing a coordinate translation $r\to r-f_0$. This coordinate transformation changes the position of the horizon but does not change the temperature of the geometry. In this system there is a conserved quantity along the radial direction $\frac{f'hn-h'fn}{\sqrt{h}}$ associated with a scaling symmetry in the background equations of motion. We can also derive this radially conserved quantity from a linear combination of the $tt$ and $zz$ components of the Einstein's equations of motion. From this conserved quantity we find that at zero temperature $n_2=M/2$, and at nonzero temperature $2 n_2=M-\pi T n_0 \sqrt{h_0}$.

In numerics, $n_0$ and $h_0$ can be fixed to arbitrary values and finally need to be rescaled according to the boundary coefficients in front of $r^2$ in $n$ and $h$. In this case, the physical value of the magnetic field $B$ will also be rescaled and become different from the input value of $B$. In our numerics we fix $n_0$ and $h_0$ to numerically convenient values for simplicity, and we can read out the physical value of the magnetic field from the boundary values of the metric fields by $B=\tilde{B} r^2/g_{xx}|_{r\to \infty}$.
With the two free parameters $T$, $\tilde{B}$ at the horizon, we can integrate the equations to the boundary and produce background solutions characterized by the temperature $T$ and the physical magnetic field $B$.

\subsection{Longitudinal magnetoconductivity}

To calculate the longitudinal magnetoconductivity in the backreacted geometry above, we consider perturbations $~\delta V_z=v_z e^{-i\omega t},~\delta A_t=a_t e^{-i\omega t}$ on the background solutions. As we are studying the system at zero charge and axial charge density, these perturbations do not couple to the metric perturbations. The equations for $v_z$ and $a_t$ are
\bea\label{eqnsvzat}
a_t'+\frac{8 \alpha B}{n\sqrt{h}}v_z&=&0\,,\\
v_z''+\Big(\frac{f'}{f}+\frac{n'}{n}-\frac{h'}{2h}\Big)v_z'+\frac{\omega^2}{f^2}v_z+\frac{8\alpha B\sqrt{h}}{nf}a_t'&=&0\,,
\eea
and we can simplify them into one single equation for $v_z$
\be
v_z''+\Big(\frac{f'}{f}+\frac{n'}{n}-\frac{h'}{2h}\Big)v_z'+\Big(\frac{\omega^2}{f^2}-\frac{(8\alpha B)^2}{n^2f}\Big)v_z=0.
\ee

At the boundary, the asymptotic behavior for $v_z$ is\be v_z\simeq v_{z0}\Big{(}1+ \frac{\omega^2\ln r}{2r^2}\Big{)}+\frac{v_{z1}}{r^2}\ee and the definition of the conductivity is \cite{Horowitz:2008bn} \be\label{sigmadef}\sigma=\frac{2 v_{z1}}{i \omega v_{z0}}+ \frac{i \omega}{2}\ee under infalling boundary conditions at the horizon. This definition of conductivity corresponds to the retarded two point function of the consistent current with the covariant current, whose definition can be found in \cite{{Landsteiner:2012kd},Landsteiner:2014vua}.

Without an exact background solution, it is not possible to solve this equation analytically. We instead solve it numerically by integrating the equation from the horizon to the boundary with infalling boundary condition at the horizon. As shown in \cite{Landsteiner:2014vua}, in general for a chiral anomalous system we need to impose three kinds of dissipations in order to make the DC magnetoconductivity finite, including the energy, momentum and axial charge dissipations. Here as a special case of zero charge and axial charge density, only the axial charge dissipation is needed for a finite DC magnetoconductivity, which we will consider in the next section. In this section, with no axial charge dissipation mechanism, the imaginary part of the longitudinal magnetoconductivity behaves as $1/\omega$ at $\omega\to 0$ and the real part consequently gets a $\delta$-function at $\omega=0$, which means that the DC magnetoconductivity is divergent. The longitudinal magnetoconductivity takes the form of \be \sigma_{zz}=\sigma_{E}+\frac{i}{w}c_0\ee at low frequency, where $\sigma_{E}$ is the quantum critical conductivity.

Different from the probe limit, the backreacted background solutions depend on the value of $\lambda B$ but not $\alpha$ while the perturbations only depend on $\alpha B$ so the final result of the magnetoconductivity will depend on all three parameters of $\lambda$, $\alpha$ and $B/T^2$. As $\lambda$ or $B/T^2$ increases, the effects of backreactions will become more apparent.

\begin{figure}[h]
    \centering
    \begin{tabular}{@{}cccc@{}}
    \includegraphics[width=.47\textwidth]{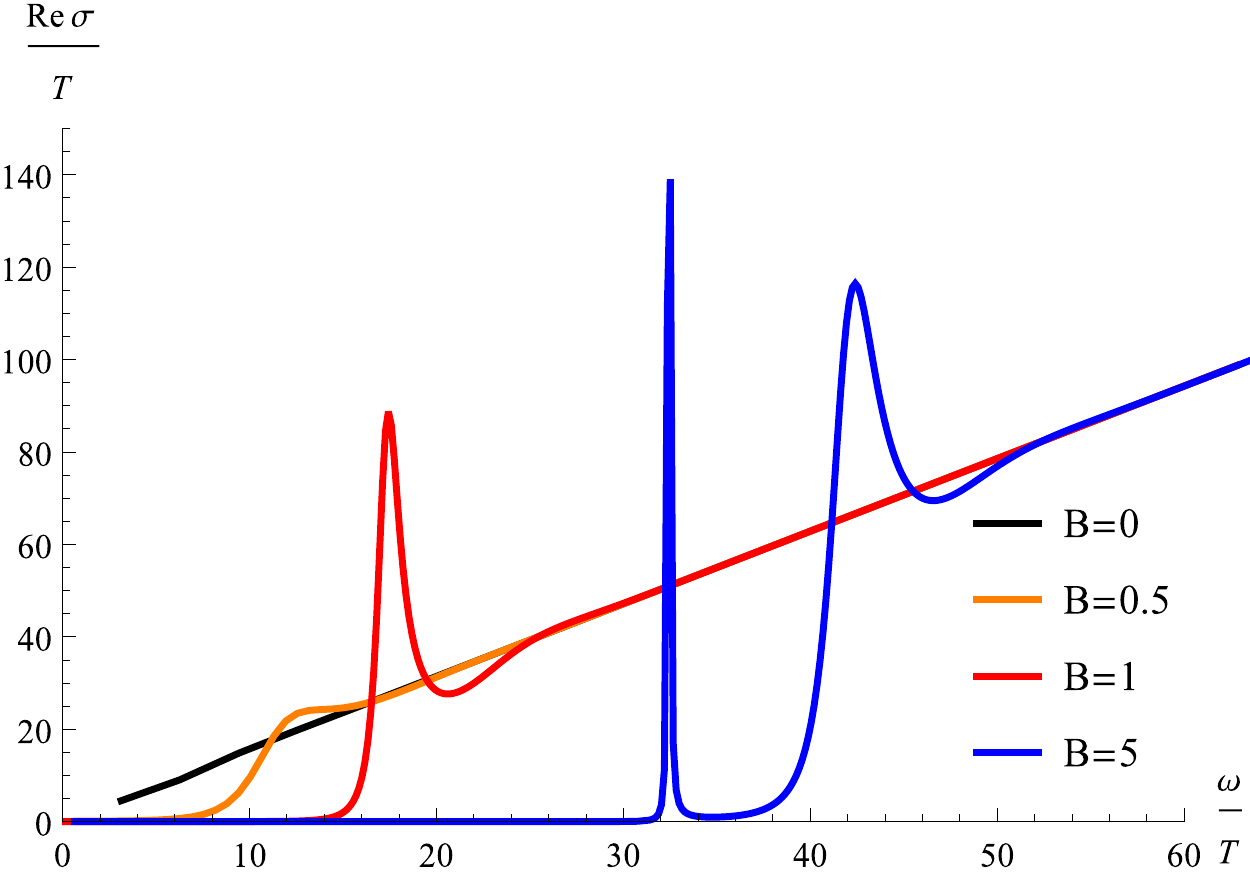} &
    \includegraphics[width=.47\textwidth]{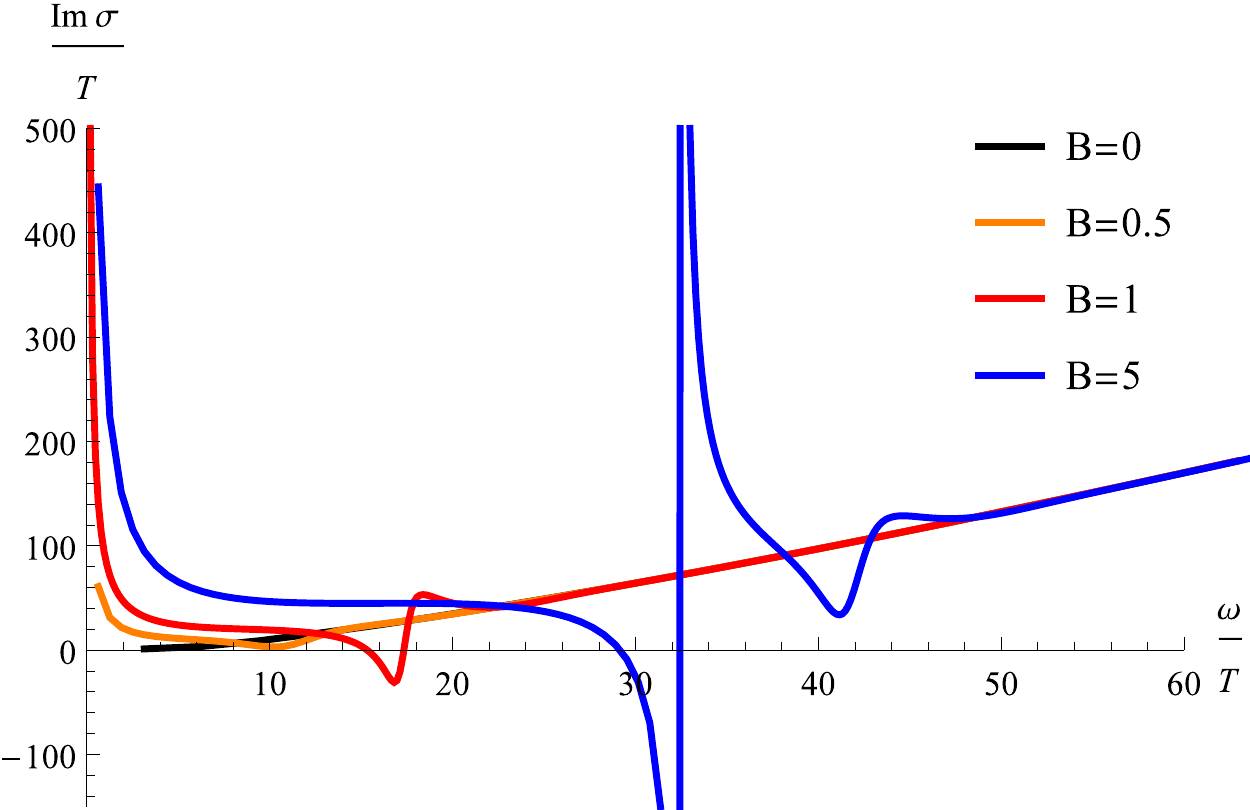}  \\
  \end{tabular}
  \caption{\small Real (left) and imaginary (right) parts of the AC longitudinal magnetoconductivity for different values of $B=0~\text{(Black)},~0.5~ \text{(Orange)},~1 ~\text{(Red)},~5$ (Blue). $\lambda=1$, $\alpha=1$ and $T=1/\pi$.}    \label{fig:ac1}
\end{figure}

In Fig.~\ref{fig:ac1}, we show the real part and the imaginary part of the AC magnetoconductivity as a function of $\omega/T$ for different values of $B/T^2$ separately at fixed $\lambda=1$ and $\alpha=1$. The $\delta$-function at $\omega=0$ cannot be seen from the real part of this figure but as we will see from the coefficient in front of $1/\omega$ in the imaginary part, the height of the $\delta$-function grows as a function of $B/T^2$. As $B$ increases, the gap region in the real part becomes wider and wider as can be seen from the figure, which is consistent with the fact that weight is transferred to the $\omega\to 0$ region as $B$ increases. At larger values of $\omega/T$ quasinormal modes start to show up which lead to peaks in the real part of $\sigma_{zz}$ and as $B$ increases, more and higher peaks will arise. This behavior was also found in the axial charge dissipation system in the probe limit \cite{Jimenez-Alba:2014iia}.

\begin{figure}[h]
    \centering
    \begin{tabular}{@{}cccc@{}}
    \includegraphics[width=.47\textwidth]{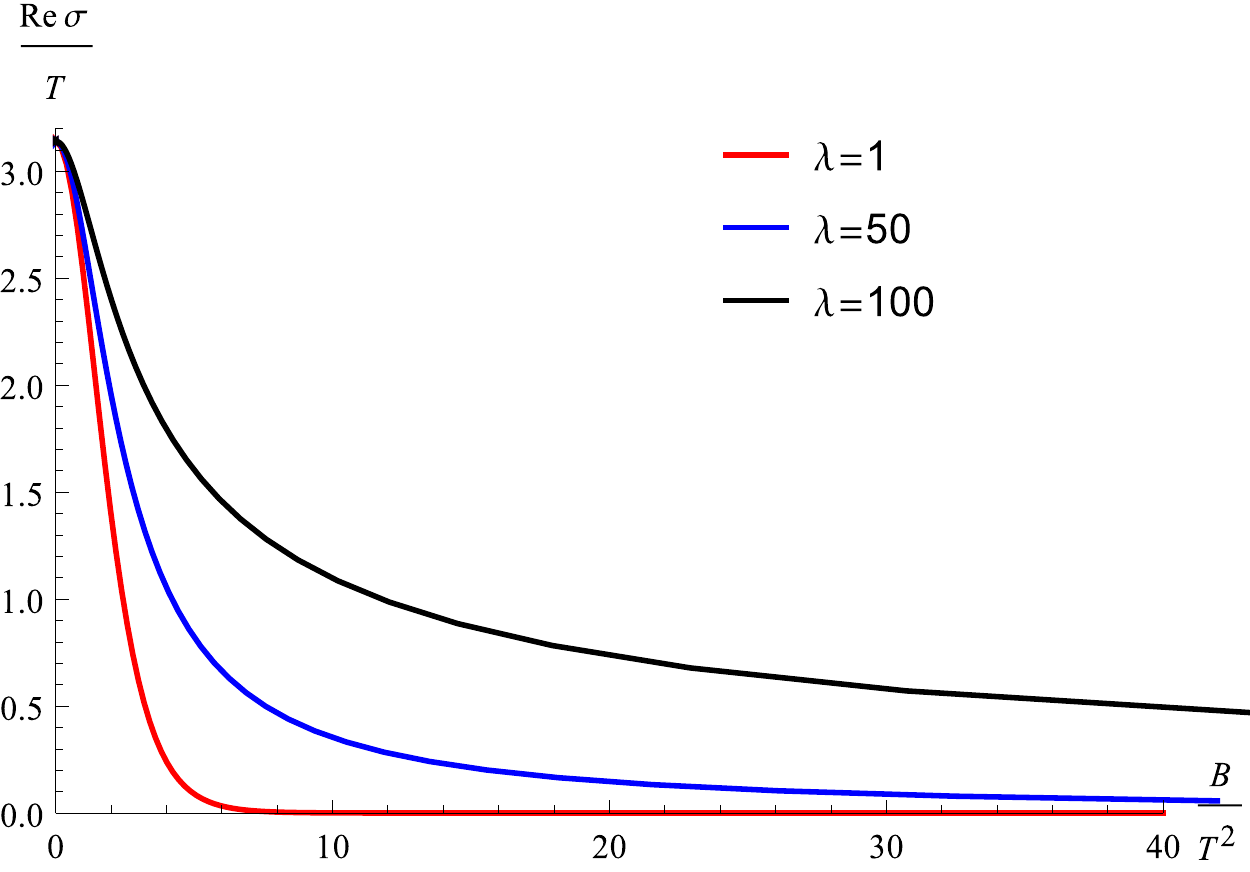} &
    \includegraphics[width=.47\textwidth]{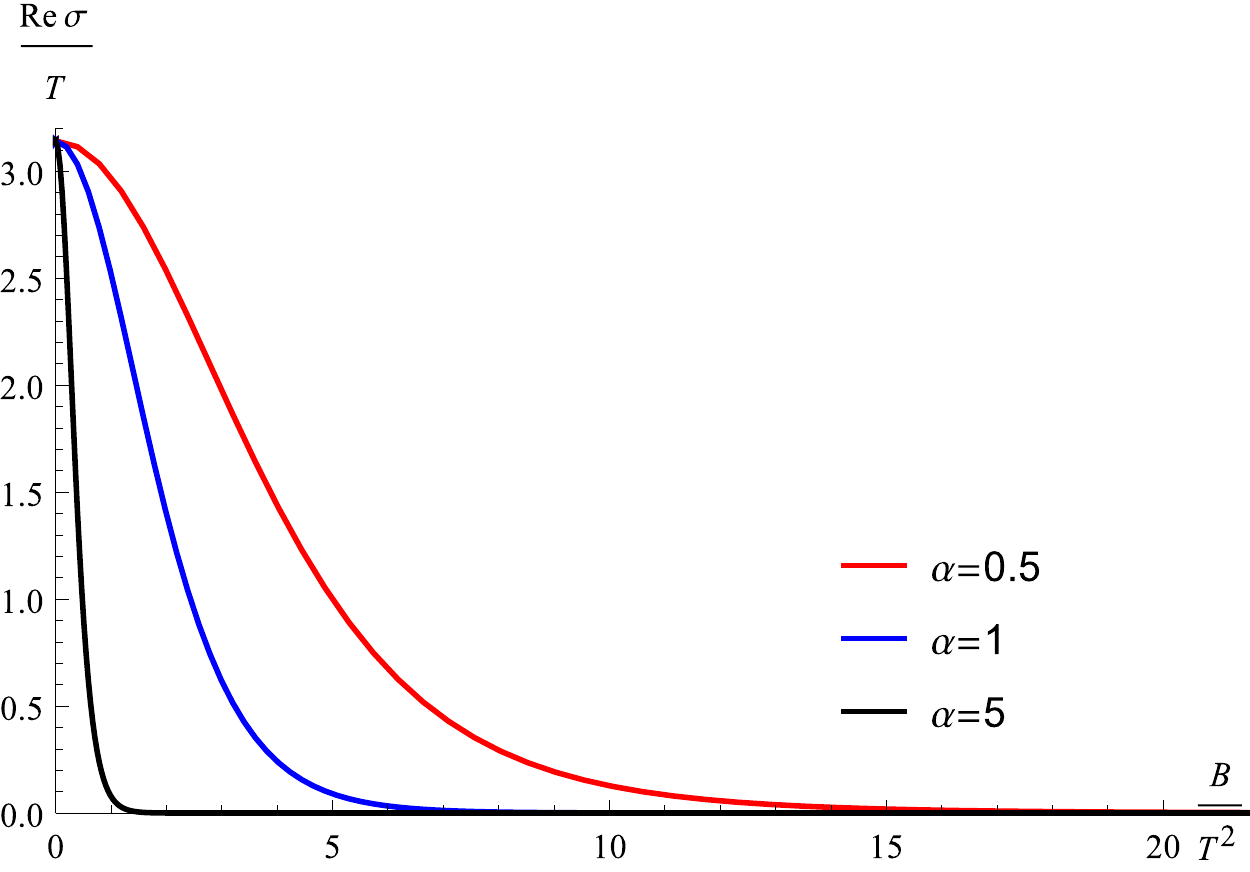} \\
  \end{tabular}
  \caption{\small Left: real part of the DC longitudinal magnetoconductivity as a function of $B/T^2$ for $\alpha=1$ and $\lambda=1,~50,~100$. Right:  real part of the DC longitudinal magnetoconductivity as a function of $B/T^2$ for $\lambda=1$ and $\alpha=0.5,~1,~5$.}    \label{fig:dcre1}
\end{figure}

In Fig.~\ref{fig:dcre1}, we show the real part of the DC magnetoconductivity (excluding the $\delta$-function), i.e. the quantum critical conductivity $\sigma_{E}$, as a function of $B/T^2$ for various values of $\lambda$ and $\alpha$. In the left figure, we fix $\alpha=1$ and choose $\lambda=1,~50,~100$ respectively. In the right figure, we fix $\lambda=1$ and choose $\alpha=0.5,~1,~5$ respectively.   When $B=0$, $\text{Re} \sigma_{zz}(0)=\pi T$, which is universal regardless of the value of $\alpha$ or $\lambda$. From the left figure in Fig.~\ref{fig:dcre1}, we can see that at $\alpha=1$, when $\lambda$ becomes larger, $\text{Re} \sigma_{zz}(0)$ decreases more slowly with increasing $B$, which deviates from the probe case, but the qualitative behavior is the same and as we will see later $\text{Re} \sigma_{zz}(0)$ finally vanishes at $B\to \infty$ as long as $\lambda$ is not too large.  From the right figure we can see that at $\lambda=1$, $\text{Re} \sigma_{zz}(0)$ decreases monotonically as $B$ increases at values of $\alpha$ not too small.

\begin{figure}[h]
    \centering
    \begin{tabular}{@{}cccc@{}}
    \includegraphics[width=.47\textwidth]{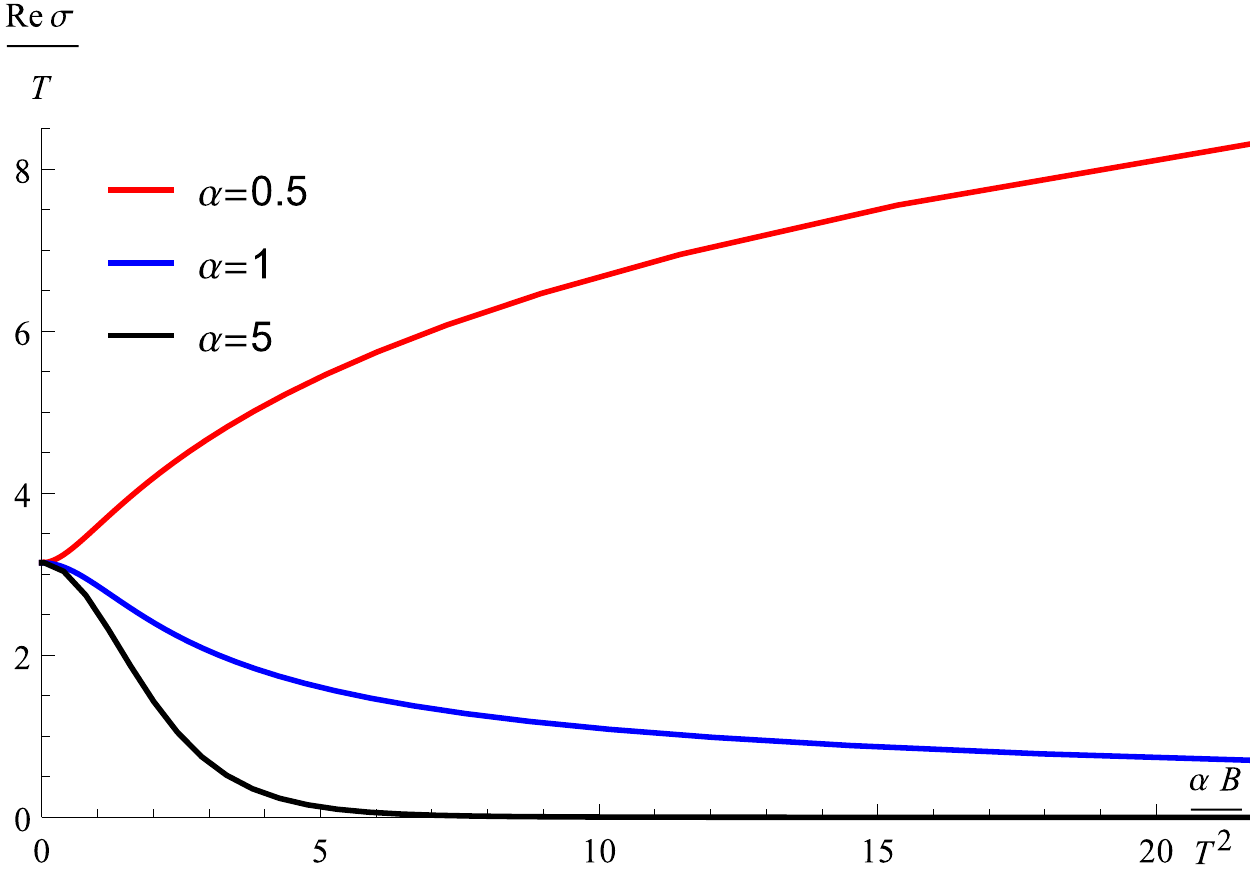} &
    \includegraphics[width=.47\textwidth]{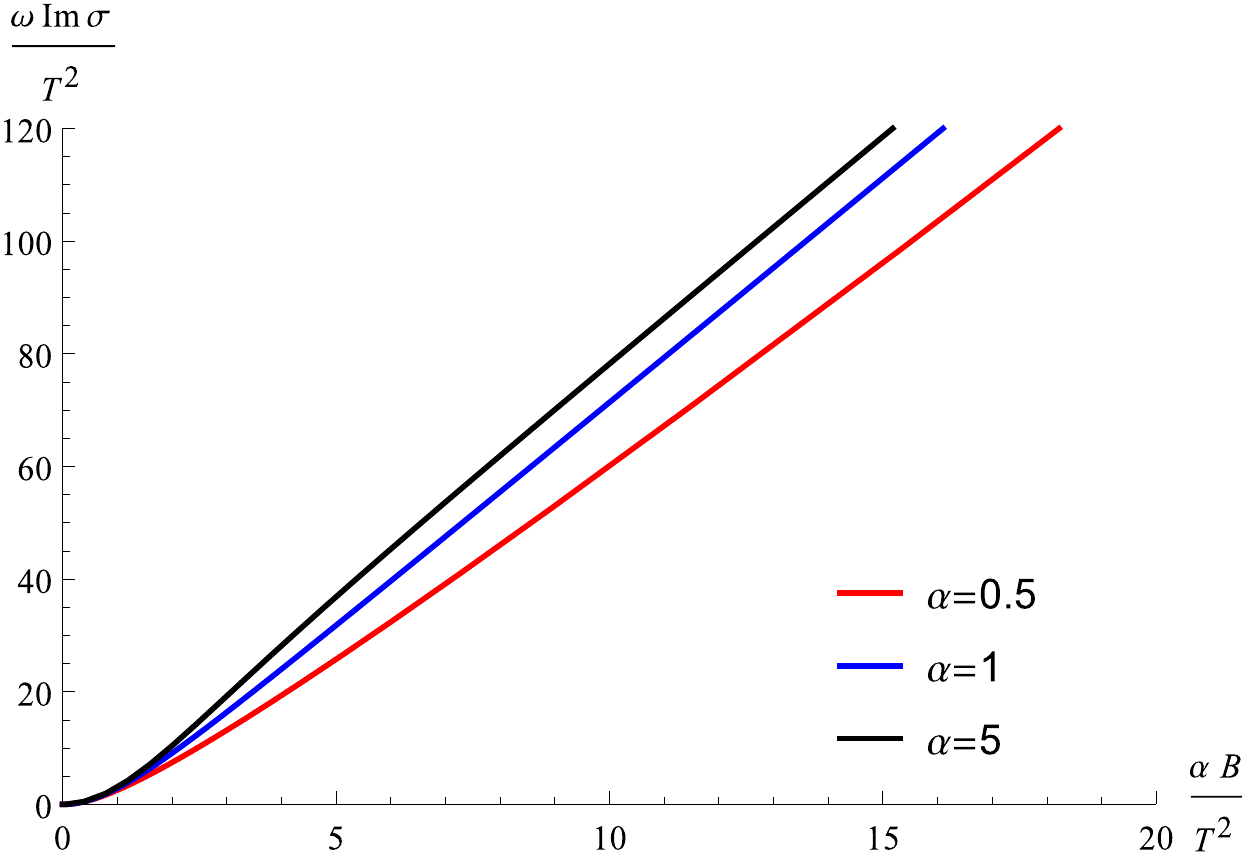} \\
  \end{tabular}
  \caption{\small Real (left) and imaginary (right) parts of the DC longitudinal magnetoconductivity at large $\lambda=100$ for $\alpha=0.5,~1,~5$ respectively.}    \label{fig:dclargelambda}
\end{figure}

However, when $
\lambda$ is very large while $\alpha$ very small, $\text{Re} \sigma_{zz}(0)$ would start to grow monotonically as $B$ increases. Fig.~\ref{fig:dclargelambda} shows that this would be the case for $\lambda=100$ and $\alpha=0.5$. This behavior is related to the divergence of $\text{Re} \sigma_{zz}(\omega\to 0)/T$ for small $\alpha<\sqrt{\lambda}/16$ at the limit $B/T^2 \to \infty$ as we will explain in the next subsection for the zero temperature limit. Note that in this figure, the horizontal axis is $\alpha B/T^2$ instead of $B/T^2$, and this shows that $\sigma_{zz}(0)$ depends on both $\alpha$ and $B/T^2$ separately.

\begin{figure}[h]
    \centering
    \begin{tabular}{@{}cccc@{}}
    \includegraphics[width=.47\textwidth]{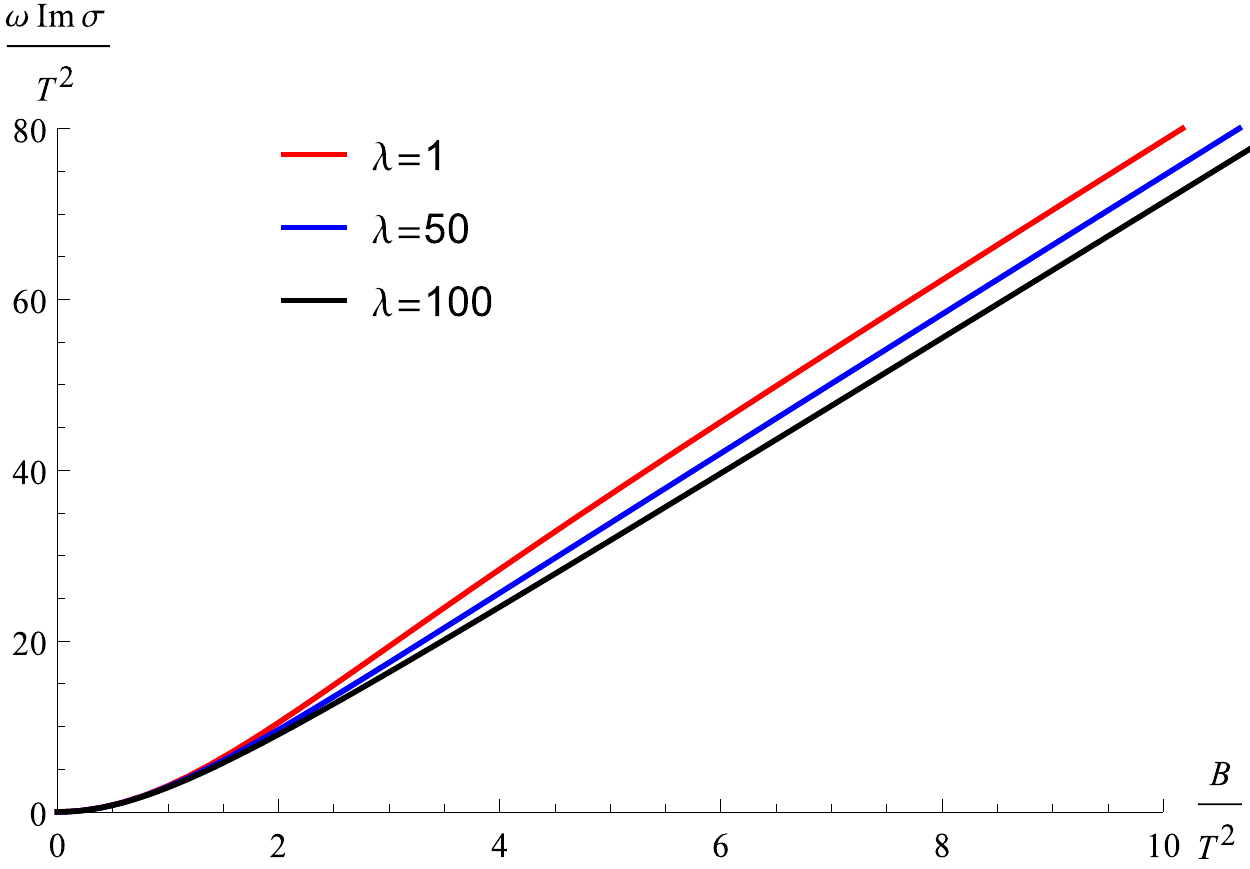} &
    \includegraphics[width=.47\textwidth]{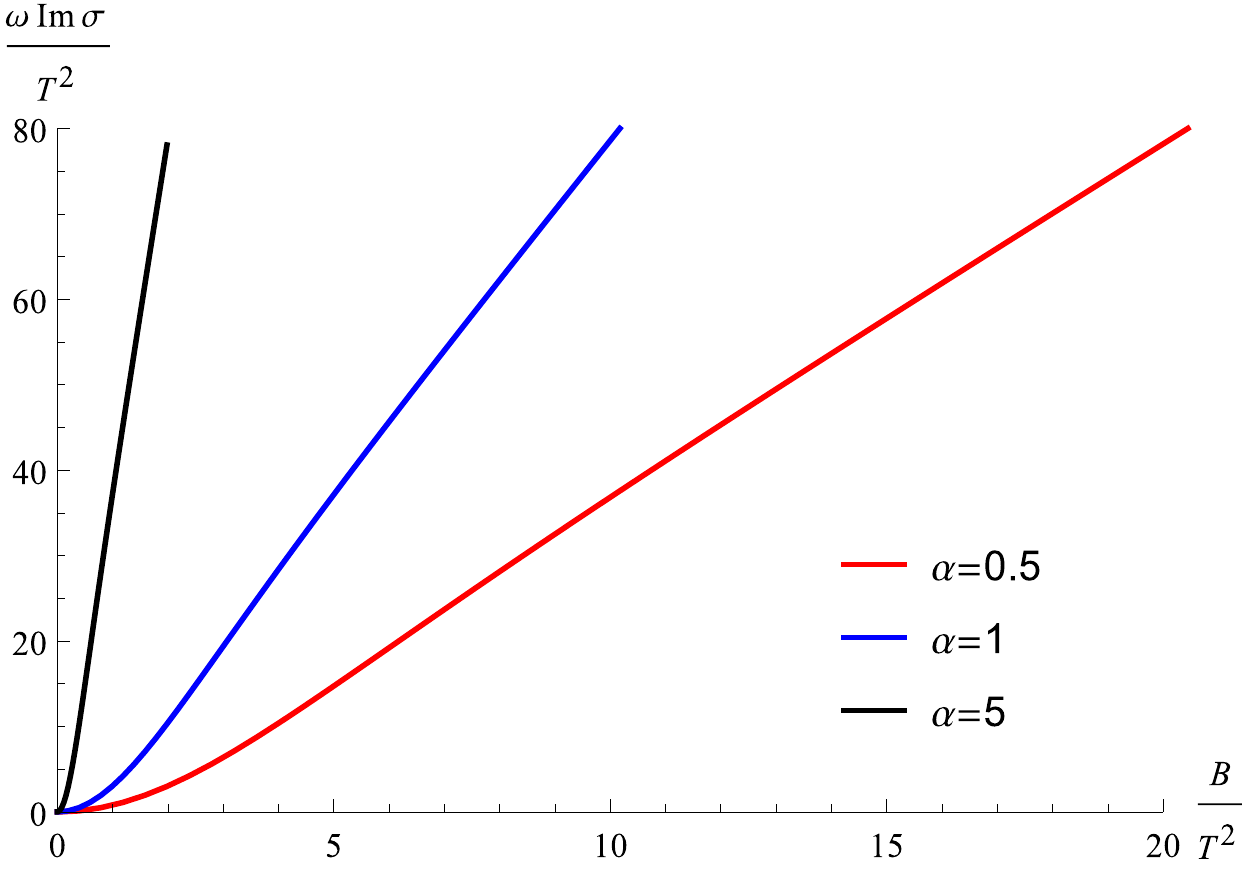} \\
  \end{tabular}
  \caption{\small Left: coefficient in front of $1/\omega$ in the imaginary part of the DC longitudinal magnetoconductivity as a function of $B/T^2$ for $\alpha=1$ and $\lambda=1,~50,~100$. Right: coefficient in front of $1/\omega$ in the imaginary part of the DC longitudinal magnetoconductivity as a function of $B/T^2$ for $\lambda=1$ and $\alpha=0.5,~1,~5$.}    \label{fig:dcim1}
\end{figure}

\begin{figure}[h]
    \centering
    \includegraphics[width=.65\textwidth]{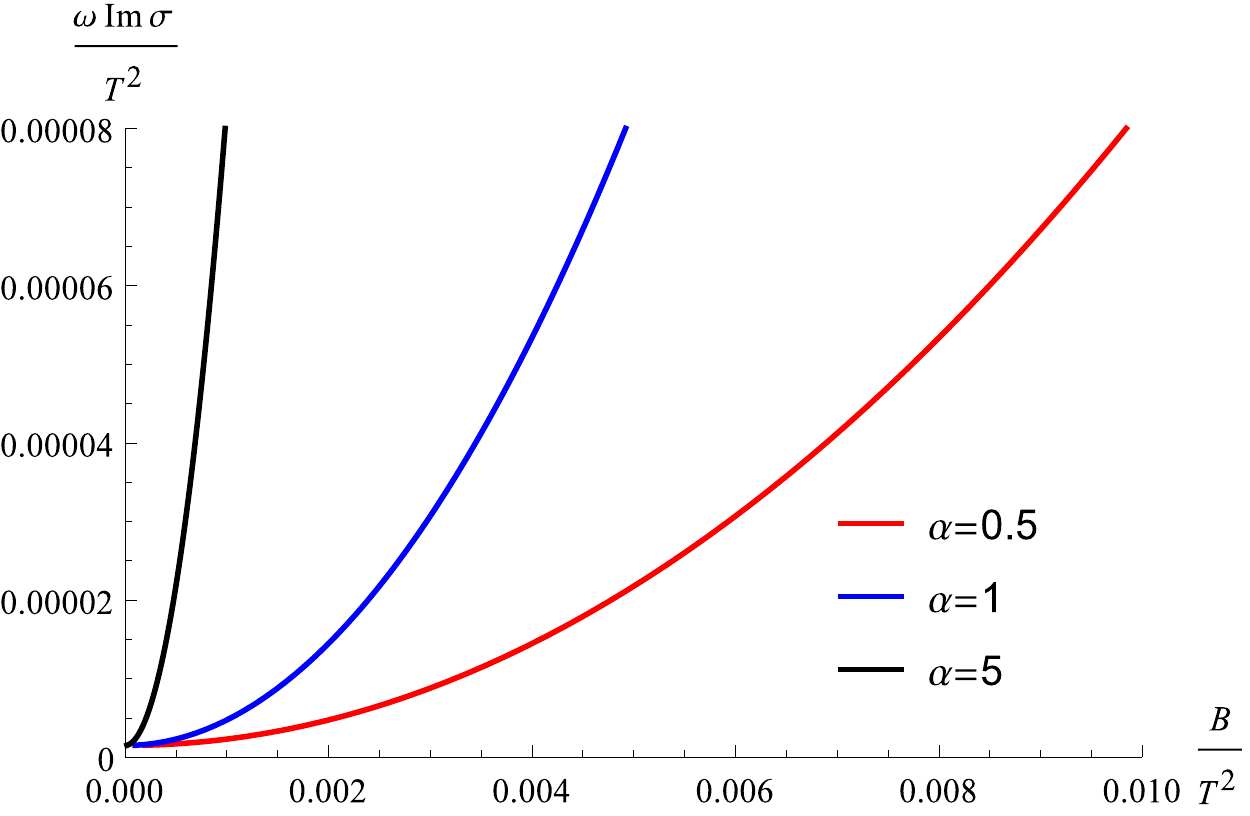}
  \caption{\small Small $B/T^2$ region for the coefficient in front of $1/\omega$ in the imaginary part of the DC longitudinal magnetoconductivity for $\lambda=1$ and $\alpha=0.5,~1,~5$.}    \label{fig:dcim2}
\end{figure}

In Fig.~\ref{fig:dcim1}, we show the imaginary part of the DC magnetoconductivity $\text{Im} \sigma_{zz}(\omega\to 0)$ as a function of $B$ for various values of $\lambda$  and $\alpha$. Because there is no dissipation and $\text{Im} \sigma_{zz}(0)$ behaves as $1/\omega$ near $\omega\to 0$, we plot the coefficient in front of $1/\omega$ in $\text{Im} \sigma_{zz}(0)$ in the figure. When $\lambda$ increases the deviation from the probe limit becomes more apparent, but similar to the real part, the qualitative behavior is still the same as the probe limit. In Fig.~\ref{fig:dcim2}, we zoom in at the small $B/T^2$ region. At small $B/T^2$, as shown in Fig.~\ref{fig:dcim2}, $\omega \text{Im} \sigma_{zz}(\omega\to 0)$ is proportional to $B^2$ and the coefficient does not depend on $\lambda$ as we will see from the hydrodynamic formula below, which means that it is the same as in the probe case at leading order. At large $B/T^2$, $\omega \text{Im} \sigma_{zz}(\omega\to 0)$ is linear in $B$. This result is qualitatively the same as in the probe case, however, the quantitative difference due to backreaction becomes apparent for large $B/T^2$ and large $\lambda$. To investigate the large $B$ region more carefully we will study the system at zero temperature which corresponds to the $B/T^2\to \infty$ limit in the next subsection. Surprisingly we will see that at $B/T^2\to \infty$ the behavior of $\sigma_{zz}(\omega\to 0)$ goes back to the probe result. From Fig.~\ref{fig:dclargelambda} we can see that at large $\lambda$ and small $\alpha$ when the real part starts to diverge at $B/T^2\to \infty$, the leading order in $\omega$ behavior of the imaginary part remains qualitatively the same as those with other values of $\lambda$ and $\alpha$.

As shown in \cite{Landsteiner:2014vua} the hydrodynamic formula for the small $\omega$ longitudinal magnetoconductivity at both zero charge and axial charge density, which the holographic probe system obeys is
\be
\sigma_{zz}=\sigma_{E}+\frac{i}{\omega}\frac{(8\alpha B)^2}{\chi_5},
\ee where $\sigma_E$ is the quantum critical conductivity, $\chi_5=\partial \rho_5/\partial\mu_5$ is the static axial charge susceptibility. To calculate $\chi_5$, we start from the following equation for $a_t$ which can be obtained from equations (\ref{eqnsvzat}) for $\omega\to 0$
\be\label{eqnforat}
(n\sqrt{h}a_t')'=\frac{(8\alpha B)^2\sqrt{h}}{n f}a_t.
\ee Here an integration constant from the equation of motion for $v_z$ has been chosen to be zero from the boundary conditions at the horizon. At finite temperature and small $B/T^2$, we can solve this equation order by order in $B/T^2$ and the leading order contribution to $\chi_5$ only depends on $B$ from the background small $\lambda B/T^2$ corrections, which means that at small $B/T^2$ and small $\lambda B/T^2$, the leading order contribution to $\omega\text{Im}\sigma_{zz}(0)$ is $\frac{(8\alpha B)^2}{2\pi^2 T^2}$, which is the same as the probe limit and is subject to $\lambda B/T^2 $ and $B^2/T^4$ order corrections. This is also consistent with our numerical findings. This hydrodynamic formula should be valid in the hydrodynamic regime:$B/T^2\ll 1$, which is indeed the case as in this regime the leading order result is the same as the probe limit result. In the following subsection we will see that even at zero temperature, this hydrodynamic formula still agrees with the holographic result.

\subsection{Zero temperature}

The zero temperature limit of this system is equivalent to the large $B/T^2$ limit at finite temperature. Due to numerical difficulties at large $B/T^2$ in the finite temperature calculation, in this subsection we study the system at exact zero temperature to approach the $B/T^2\to \infty$ limit. At zero temperature, an exact solution to this system is AdS$_3\times$ R$^2$ \cite{{D'Hoker:2009mm},{D'Hoker:2010hr},{Mamo:2013efa}}. We need irrelevant perturbations at the horizon to flow this solution to asymptotic AdS$_5$ solutions. The near horizon solution with irrelevant perturbations is
\bea ds^2=-3 r^2(1+f_1 r^{\beta}&+&\cdots) dt^2+\frac{dr^2}{3  r^2(1+f_1 r^{\beta}+\cdots)}+r^2(1+f_1 r^{\beta}+\cdots)dz^2+ \nonumber\\ \frac{B\sqrt{\lambda}}{2\sqrt{3}}\bigg{(}&1&-\frac{19+2\sqrt{57}}{14}f_1 r^{\beta}+\cdots\bigg{)}(dx^2+dy^2),\eea
where the ``$\cdots$'' are higher order corrections and $f_1$ has to be negative in order to flow to asymptotic AdS$_5$ solutions. The value of $\beta$ can be solved from the equations of motion for the perburbations to be $\beta=\frac{1}{3}(\sqrt{57}-3)$. As $B$ can always be absorbed into rescaling of $x$ and $y$, and $f_1$ can also be rescaled to $-1$ by rescaling $r$, it seems that we can only get one effective value of $B$ at zero temperature. However, when we scale $r$ to rescale $f_1$ to $-1$, the boundary behavior of $g_{xx}$ will change accordingly with a different coefficient in front of $r^2$ and leads to a different physical value of $B$: the physical magnetic field for $f_1=-1$ is $(-f_1)^{2/\beta}$ times the physical magnetic field for other values of $f_1$. Thus tuning the near horizon parameter $f_1$ will give solutions with different values of $B$, though these solutions are in fact equivalent physically as $B$ is the only scale in the background solutions at zero temperature.

 From the background equations it looks like that the value of $\lambda$ only affects the details of the one to one correspondence between the horizon initial value of $f_1$ and the final value of the physical magnetic field, however, with different values of $\lambda$ the background geometry is different even for the same physical magnetic field. Thus $\lambda$, which does not appear in the equations of perturbations, would still affect the conductivity at zero temperature. The only dimensionful quantity at zero temperature is $B$, thus from dimensional analysis, we know that $\text{Re} \sigma_{zz}(0)\sim \sqrt{B}$ while $\omega \text{Im}\sigma_{zz}(0)\sim B$. Due to the fact that the background geometry depends on $\lambda$ while the equations of perturbations only depend on $\alpha$, the behavior of $\sigma_{zz}(\omega\to 0)$ is expected to depend on both $\alpha$ and $\lambda$. As we will explain below, there are three different kinds of qualitative behavior of $\sigma_{zz}(\omega\to 0)$ depending on the value of $\alpha/\sqrt{\lambda}$: for $\alpha/\sqrt{\lambda}=1/32$,  $\text{Re} \sigma_{zz}(0)/\sqrt{B}$ is a constant at zero frequency, for $\alpha/\sqrt{\lambda}>1/32$,  $\text{Re} \sigma_{zz}(0)/\sqrt{B}=0$ at leading order in $\omega$, and for $\alpha/\sqrt{\lambda}<1/32$, $\text{Re} \sigma_{zz}(0)/\sqrt{B}$ diverges as $\omega^{32 \alpha/\sqrt{\lambda}-1}$.

This characterization of the three different kinds of qualitative behaviors can be derived from the IR equations using the near far matching method \cite{Faulkner:2009wj} as follows. The equation of motion for $v_z(r)$ reduces to the following in the IR region $r\ll \sqrt{B}$
\be
v_z''+\frac{v_z'}{r}+\Big(\frac{\omega^2}{9 r^4}-\frac{(16\alpha)^2}{ \lambda r^2}\Big)v_z=0.
\ee The infalling solution to this equation is the Bessel $K$-function $
K_{\frac{16\alpha}{\sqrt{\lambda}}}\big(\frac{-i \omega}{3 r}\big)$. Expanding this function at the boundary of the IR region $\omega\ll r\ll \sqrt{B}$ we can get the relative coefficient in front of the two linearly independent solutions $v_{z}^{(1)}|_{\omega\ll r\ll \sqrt{B}}= r^{\frac{16\alpha}{\sqrt{\lambda}}}+\cdots$ and $v_{z}^{(2)}|_{\omega\ll r\ll \sqrt{B}}=r^{-\frac{16\alpha}{\sqrt{\lambda}}}+\cdots$ of this region. It turns out that the relative coefficient of the two solutions scales as $\omega^{\frac{32\alpha}{\sqrt{\lambda}}}$ with a complex coefficient. The two linearly independent solutions $r^{\frac{16\alpha}{\sqrt{\lambda}}}$ and $r^{-\frac{16\alpha}{\sqrt{\lambda}}}$ are both real so the boundary coefficients $v_{z0}^{(1,2)}$ and $v_{z1}^{(1,2)}$ associated with these two solutions are all real. Substituting these into the formula for $\sigma_{zz}$ in (\ref{sigmadef}) it is easy to see that the leading order in the imaginary part $\text{Im}\sigma_{zz}(\omega\to 0) \sim \frac{2 v_{z1}^{(1)}}{ i\omega v_{z0}^{(1)}} $ while the leading order contribution to the real part scales as $\text{Re}\sigma_{zz}(\omega\to 0) \sim \omega^{32\alpha/\sqrt{\lambda}-1}$. Thus more explicitly the scaling behavior of the small frequency longitudinal magnetoconductivity at zero temperature is the following.
\begin{itemize}
\item For $\frac{32\alpha}{\sqrt{\lambda}}<1$, $\text{Re} \sigma_{zz}(\omega\to 0)\sim c_{1}(\lambda,\alpha) \sqrt{B}\big(\frac{\omega}{\sqrt{B}}\big)^{\frac{32\alpha}{\sqrt{\lambda}}-1}$ and $\text{Im} \sigma_{zz}(\omega\to 0)\sim \frac{B}{\omega}d(\lambda,\alpha)$;
\item For $\frac{32\alpha}{\sqrt{\lambda}}=1$, $\text{Re} \sigma_{zz}(\omega\to 0)\sim c_{2}(\lambda,\alpha)\sqrt{B}$ and $\text{Im} \sigma_{zz}(\omega\to 0)\sim \frac{B}{\omega}d(\lambda,\alpha)$;
\item For $\frac{32\alpha}{\sqrt{\lambda}}>1$, $\text{Re} \sigma_{zz}(\omega\to 0)\sim 0$ and $\text{Im} \sigma_{zz}(\omega\to 0)\sim \frac{B}{\omega}d(\lambda,\alpha)$,
\end{itemize}
where $c_{1,2}(\lambda,\alpha)$ and $d(\lambda,\alpha)$ are constants which might depend on $\lambda$ and $\alpha$ while do not depend on $\omega$ or $B$. The condition that this behavior only exists for $\sqrt{\lambda}> 32\alpha$ shows that this is a backreaction effect which cannot be seen in the probe limit. This also explains the strange monotonically increasing behavior for the finite temperature $\text{Re} \sigma_{zz}(\omega\to 0)$ with $B$ for large values of $\sqrt{\lambda}$ compared to $\alpha$.

Note that to compare this result with the finite temperature case of last subsection, we should focus on the $B$ scaling instead of the $\omega$ scaling behavior because in numerics we always have a small while nonvanishing value of $\omega$. The $B$ scaling behavior for the real part of the longitudinal magnetoconductivity is \be \text{Re} \sigma_{zz}(\omega\to 0)\sim c_{1}(\lambda,\alpha) B^{1-\frac{16\alpha}{\sqrt{\lambda}}}{\omega}^{\frac{32\alpha}{\sqrt{\lambda}}-1},\ee which means that for $\frac{16\alpha}{\sqrt{\lambda}}<1$, the real part of the finite temperature DC longitudinal magnetoconductivity would diverge at $B/T^2 \to\infty$, which is consistent with the numeric result of last subsection.

We confirm this analytic finding with numerics. Numerically we obtain the zero temperature background solutions with different values of magnetic field by choosing different initial values of $f_1$ at the horizon. Then we solve the equation of motion for $v_{z}$ with infalling boundary condition at the horizon and read the boundary coefficients of $v_{z0}$ and $v_{z1}$ with the solutions for $v_{z}$.

For $\frac{32\alpha}{\sqrt{\lambda}}>1$, we numerically checked that for a continuous range of parameters $\alpha$ and $\lambda$,
$\text{Re} \sigma_{zz}(\omega\to 0)\sim 0$ and $\text{Im} \sigma_{zz}(\omega\to 0)\sim 8\alpha \frac{B}{\omega}$, which coincides with the large $B/T^2$ probe limit result at leading order in $\omega$. This is also consistent with the large $B/T^2$ behavior of the backreacted finite temperature results in this parameter region. This numerical finding shows that in the small $\lambda$ region $\sqrt{\lambda}<32\alpha$ the result for the DC longitudinal magnetoconductivity still agrees with the probe limit result quantitatively at leading order in $\omega$. However, at subleading orders of $\omega$ in both the real and imaginary parts of $\sigma_{zz}$, effects of $\lambda$ will appear.

Then we choose $\lambda=100$ and $\alpha=5/32$, which gives $32\alpha/\sqrt{\lambda}=1/2<1$. We find that $\text{Re} \sigma_{zz}(\omega\to 0)$ indeed scales as $c_{1} \sqrt{B}\big(\frac{\omega}{\sqrt{B}}\big)^{-1/2}$ where $c_1$ is around $0.84$ at $\lambda=100$ and $\alpha=5/32$. The imaginary part $\text{Im} \sigma_{zz}(\omega\to 0)\sim \frac{B}{\omega}d(\lambda,\alpha)$ where $d$ is still $8\alpha$ for this set of values of $\lambda$ and $\alpha$.

At $\lambda=100$ and $\alpha=5/16$, which gives $32\alpha/\sqrt{\lambda}=1$, we find that $\text{Re} \sigma_{zz}(\omega\to 0)$ indeed scales as $c_{2} \sqrt{B}$ where $c_2$ is around $1.21$ at $\lambda=100$ and $\alpha=5/16$. Again the imaginary part $\text{Im} \sigma_{zz}(\omega\to 0)\sim \frac{B}{\omega}d(\lambda,\alpha)$ where $d$ is $8\alpha$ for this set of values of $\lambda$ and $\alpha$. We expect that for all values of $\alpha$ and $\lambda$ the leading order in $\omega$ behavior of $\text{Im} \sigma_{zz}(\omega\to 0)$ is always $8\alpha \frac{i}{\omega}$. The zero temperature divergence of the quantum critical conductivity was also found in Einstein-dilaton systems at zero density when there is no chiral anomaly \cite{Kiritsis:2015oxa}.



We can now check if the hydrodynamic formula is still valid at zero temperature, which is already out of the hydrodynamic regime. At zero temperature the equation for $a_t$ is still the same as the finite temperature one of (\ref{eqnforat}) and we can solve it numerically on the zero temperature background. From dimensional analysis $\chi_5 \sim B$, and numerically we find that for any value of $\lambda$, which is larger or smaller or equal to $(32\alpha)^2$, we always have $\chi_5=8\alpha B$. By substituting $\chi_5$ into the hydrodynamic formula we find that this formula still gives the exact holographic result at leading order in $\omega$ even at $T=0$ which is outside the hydrodynamic regime. The fact that $\chi_5=8\alpha B$ for all values of $\alpha$ and $\lambda$ is also consistent with that the imaginary part of $\sigma_{zz}$ is always $8\alpha B/\omega$ at leading order in $\omega$. At the same time, the explicit value of the quantum critical conductivity cannot be obtained from the hydrodynamic formula.

The results of this subsection show that in holography we can find a parameter region in which the real part of the longitudinal DC magnetoconductivity, i.e. the quantum critical conductivity diverges at $B/T^2\to\infty$, in contrast to the previous probe limit result where the quantum critical conductivity always vanishes at $B/T^2\to\infty$.

\section{Adding axial charge dissipation}

In this section, we add axial charge dissipation to the backreacted zero density system of last section to get a finite DC longitudinal magnetoconductivity. As shown in \cite{Jimenez-Alba:2015awa}, there are two simple mechanisms to encode axial charge dissipation: one is to introduce a mass for the $U_A(1)$ gauge field and the other is to source the system by an axially charged scalar field. However, for the massive  $U_A(1)$ gauge field case, there exists a problem that the scaling dimension of the axial current has changed, so in this section we use the second way to introduce the axial charge dissipation. The massive scalar corresponds to a massive operator which can be interpreted as the mass of
the dual fermions. We will consider the following action\footnote{We have set the curvature scale $L=1$.}
\bea
S&=&\int d^5x \sqrt{-g}\bigg[\frac{1}{2\kappa^2}\Big(R+12\Big)-\frac{1}{4 e^2}\mathcal{F}^2-\frac{1}{4 e^2}F^2+\frac{\alpha}{3}\epsilon^{\mu\nu\rho\sigma\tau}A_\mu \Big(F_{\nu\rho} F_{\sigma\tau}+3 \mathcal{F}_{\nu\rho} \mathcal{F}_{\sigma\tau}\Big)\nonumber\\&&~~~-(D_\mu\Phi)^*(D^\mu\Phi)-m^2\Phi^*\Phi\bigg]\,
\eea
with
\be
\mathcal{F}_{\mu\nu}=\partial_\mu V_\nu-\partial_\nu V_\mu\,,
~~F_{\mu\nu}=\partial_\mu A_\nu-\partial_\nu A_\mu\,,~~D_\mu=\nabla_\mu-iqA_\mu\,
\ee
where the gauge fields $V_\mu$ and $A_\mu$ correspond to the vector and axial $U(1)$ currents respectively and $\Phi$ is a complex scalar field with mass $m$. As in \cite{Jimenez-Alba:2015awa}, we choose $m^2=-3$ throughout this paper to match the dimension of the dual massive operator with the dimension of the weak coupling limit.

The equations of motion are
\bea\label{eq:eomtwou1-1}
R_{\mu\nu}-\frac{1}{2}g_{\mu\nu}\Big(R-12 -\frac{\kappa^2}{2 e^2}(\mathcal{F}^2+F^2)-(D_\mu\Phi)^*(D^\mu\Phi)-m^2\Phi^*\Phi&\Big)& \nonumber \\
-\frac{\kappa^2}{e^2} \mathcal{F}_{\mu\rho}\mathcal{F}_{\nu}^{~\rho}-\frac{\kappa^2}{e^2} F_{\mu\rho}F_{\nu}^{~\rho}-\kappa^2 ((D_{\mu} \Phi)^* D_{\nu} \Phi+(D_{\nu} \Phi)^*D_{\mu} \Phi)&=&0\\
\nabla_\nu \mathcal{F}^{\nu\mu}+2\alpha\epsilon^{\mu\tau\beta\rho\sigma} F_{\tau\beta}\mathcal{F}_{\rho\sigma}&=&0\,,\\
\label{eq:eomtwou1-2}\nabla_\nu F^{\nu\mu}+\alpha\epsilon^{\mu\tau\beta\rho\sigma} \big(F_{\tau\beta}F_{\rho\sigma}
+\mathcal{F}_{\tau\beta}\mathcal{F}_{\rho\sigma}\big)+i q\big(\Phi (D^\mu\Phi)^*-\Phi^*(D^\mu\Phi)\big)&=&0\,,\\
\label{eq:eomtwou1-3}D_\mu D^\mu\Phi-m^2\Phi&=&0\,.
\eea

\subsection{Background solutions at finite temperature}

We solve this system at finite temperature with a finite magnetic field $B$ at zero charge and axial charge density. The assumption for the background solutions is
\be
ds^2=-f(r) dt^2+\frac{d r^2}{f(r)}+n(r) (dx^2+dy^2)+h(r) dz^2,
\ee and \be V_{\mu}=(0,~0,~B y,~0,~0),~~
A_{\mu}=(0,~0,~0,~0,~0),~~\Phi=\phi(r). \ee The equations become
\bea
\frac{f'h'}{2 f h}+\frac{f' n'}{f n}+\frac{h' n'}{h n}+\frac{n'^2}{2 n^2}-\lambda {\phi'}^2-\frac{12}{f}+\frac{\lambda B^2}{2 f n^2}+\frac{\lambda m^2 \phi^2}{f}=&0&,\\
\frac{f''}{f}-\frac{ n''}{n}+\frac{h'}{2h}\bigg(\frac{f'}{f}-\frac{n'}{n}\bigg)-\frac{\lambda B^2 }{ f n^2}=&0&,\\
\frac{f''}{2f}+\frac{n''}{n}+\frac{n'}{n}\bigg(\frac{f'}{f}-\frac{n'}{4 n}\bigg)-\frac{6}{f}+\frac{\lambda B^2}{4  f n^2}+\frac{\lambda m^2 \phi^2}{2 f}+\frac{\lambda {\phi'}^2}{2}=&0&,\\ \phi''+\phi'\bigg(\frac{f'}{f}+\frac{n'}{n}+\frac{h'}{2 h}\bigg)-\frac{m^2}{f}\phi=&0&,
\eea
where $\lambda=2\kappa^2/e^2$ and we have rescaled $e \phi \to \phi$.
It is difficult to find exact finite temperature solutions to this system, so we numerically integrate the equations to produce background solutions. The leading order near horizon expansion of the fields are
\be
f\simeq 4\pi T(r-1)(1+\cdots),~~n\simeq n_0(1+\cdots),~~h\simeq h_0(1+\cdots),
\ee and
\be
\phi\simeq \phi_0(1+\cdots),
\ee where the ``$\cdots$'' denotes higher order corrections which can be solved order by order given the leading order parameters. The horizon radius can always be rescaled to $r_0=1$. The free parameters are the temperature $T$, the effective physical magnetic field related to $n_0$ or the input value $\tilde{B}$ and the initial value $\phi_0$ which is related to the boundary value of $\phi$.

At the asymptotic $AdS_5$ boundary the leading order expansions of the fields are
\bea \phi&\simeq& \frac{M}{r}\bigg(1-\frac{f_0}{r}-\bigg(1+\frac{3\eta}{2\lambda }\bigg)\frac{\lambda M^2}{3}\frac{\ln r}{r^2}+\frac{f_0^2}{r^2}+f_0\lambda M^2\bigg(1+\frac{3\eta}{2\lambda }\bigg)\frac{\ln r}{r^3}\bigg)+\frac{\psi_+}{r^3}+\cdots,\\
f&\simeq& r^2\bigg(1+\frac{2 f_0}{r}-\frac{\lambda M^2-3 f_0^2}{3 r^2}+\Big(\frac{\lambda^2 M^4}{9}-\frac{\lambda B^2 }{6 }+\frac{\eta \lambda M^4}{6}\Big)\frac{\ln r}{r^4}+\cdots\bigg),\\
n&\simeq& r^2\bigg(1+\frac{2 f_0}{r}-\frac{\lambda M^2-3 f_0^2}{3 r^2}+\Big(\frac{\lambda^2 M^4}{9}+\frac{\lambda B^2}{12 }+\frac{\eta \lambda M^4}{6}\Big)\frac{\ln r}{r^4}+\cdots\bigg),\\
h&\simeq& r^2\bigg(1+\frac{2 f_0}{r}-\frac{\lambda M^2-3 f_0^2}{3r^2}+\Big(\frac{\lambda^2 M^4}{9}-\frac{\lambda B^2}{6}+\frac{\eta \lambda M^4}{6}\Big)\frac{\ln r}{r^4}+\cdots\bigg),
\eea where $M$ corresponds to the source of the axial charged scalar field $\phi$ and $\psi_{+}$ gives the response to the source. The parameter $f_0$ can be set to zero by a coordinate transformation $r\to r-f_0$, which does not change the temperature. With the horizon parameters $T$, $n_0$ and $\phi_0$, we can integrate the system to the boundary to get solutions at temperature $T$, with magnetic field $B$ and scalar source $M$.



\subsection{Longitudinal magnetoconductivity}
To calculate the longitudinal magnetoconductivity we consider perturbations $\delta V_z=v_z(r) e^{-i \omega t}$, ~~$\delta A_t=a_t(r) e^{-i \omega t}$,~~$\delta A_r=a_r(r) e^{-i\omega t}$, ~~$\delta \Phi=\Phi_1(r,t)+i\Phi_2(r,t)=\phi_1(r) e^{-i \omega t}+i \phi_2(r) e^{-i \omega t} $ on the background above, where $\phi_1$ decouples from other modes. As discussed extensively in \cite{Jimenez-Alba:2015awa}, there are two kinds of gauge choices we can choose, $a_r=0$ or $\phi_2=0$, based on the fact that the equations in the bulk are invariant under the transformation \be\delta A_{\mu}\to \delta A_{\mu}+\partial_{\mu}\Lambda,~~\Phi_2\to \Phi_2+q\Lambda\phi.\ee

With the gauge $a_r=0$, the equations for these perturbations are
\be\label{phi2eqns}
v_z''+v_z'\bigg(\frac{f'}{f}+\frac{n'}{n}-\frac{h'}{2 h}\bigg)+\frac{\omega^2 v_z}{f^2}+\frac{8\alpha  B\sqrt{h}a_t'}{f n}=0,
\ee
\be
a_t''+a_t'\bigg(\frac{n'}{n}+\frac{h'}{2 h}\bigg)+\frac{8\alpha B v_z'}{n \sqrt{h}}-\frac{2 q^2 a_t \phi^2}{f}-\frac{2 i  q \omega \phi_2\phi}{f}=0,
\ee
and the equation for $a_r$ is
\be
-\frac{8 i \alpha B \omega v_z}{f^2 n\sqrt{h}}-\frac{i \omega a_t'}{f^2}-\frac{2  q \phi_2\phi'}{f}+\frac{2  q \phi \phi_2'}{f}=0.
\ee

At the horizon, we have the ingoing boundary conditions \bea v_z&\simeq&(r-r_0)^{-\frac{i\omega}{4\pi T}}\Bigg(v_{(0)}+\bigg(\frac{v_{(0)}\big(5 B^2 \lambda + 2n_0^2 (-12 + \lambda m_s^2 \phi_{(0)}^2)\big) (-2\pi i T + \omega) \omega}{192\pi^2 T^2 n_0^2 (2\pi T - i\omega)}\nonumber \\ &-&\frac{\alpha B a_{(1)} \sqrt{h_0} (4\pi T - i\omega)}{\pi T n_0 (2\pi T -  i\omega)}\bigg)(r-r_0)+...\Bigg)
\\ a_t&\simeq&(r-r_0)^{-\frac{i\omega}{4\pi T}}\Bigg(a_{(1)}(r-r_0)+...\Bigg)
\\ \phi_2&\simeq&(r-r_0)^{-\frac{i\omega}{4\pi T}}\Bigg(-\frac{32 v_{(0)} \alpha B \pi T+a_{(1)} \sqrt{h_0}n_0 (4\pi T - i \omega)}{8 \sqrt{h_0} \pi T n_0 \phi_{(0)} q}+...\Bigg),\eea where $v_{(0)}$ and $a_{(1)}$ are two arbitrary constants.

At the boundary we have the following expansions \bea v_z &\simeq& v_{z0} \bigg(1+\frac{\omega^2}{2}\frac{\ln r}{r^2}\bigg)+\frac{v_{z1}}{r^2}+\cdots \\ a_t &\simeq& a_{t0}-(a_{t0}M^2 q^2+i M \phi_{20}q \omega)\frac{\ln r}{r^2} +\frac{a_{t1}}{r^2}+\cdots\\ \phi_2 &\simeq& \frac{\phi_{20}}{r}\Big(1-\frac{f_0}{r}\Big)+\Big(-\frac{1}{3}\lambda M^2\phi_{20}+\frac{1}{2} \omega \big(-i a_{t0}Mq+\phi_{20} \omega \big)\Big)  \frac{\ln r}{r^3}+\frac{\phi_{21}}{r^3}+\cdots.\eea

\begin{figure}[h]
    \centering
    \begin{tabular}{@{}cccc@{}}
    \includegraphics[width=.47\textwidth]{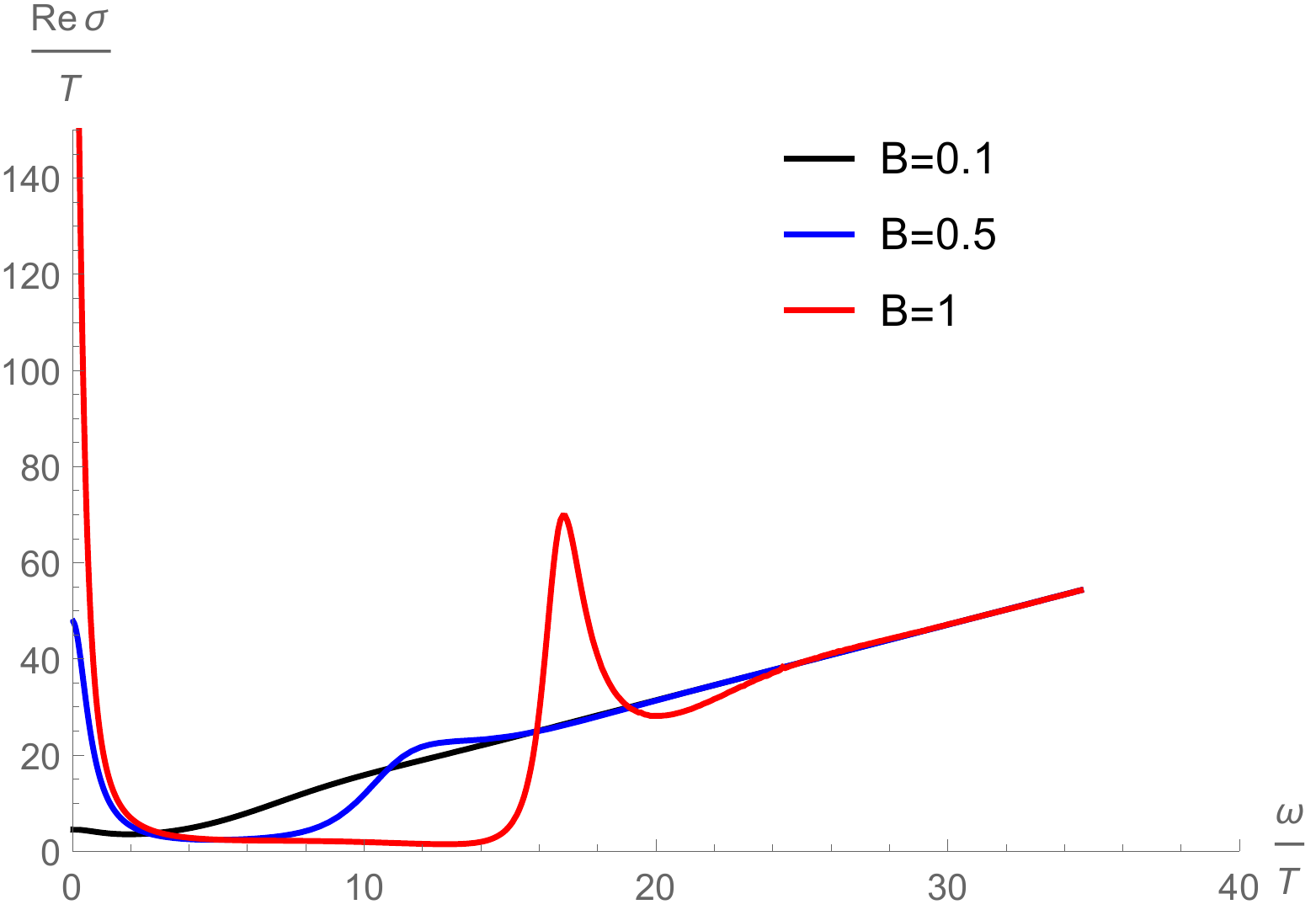} &
    \includegraphics[width=.47\textwidth]{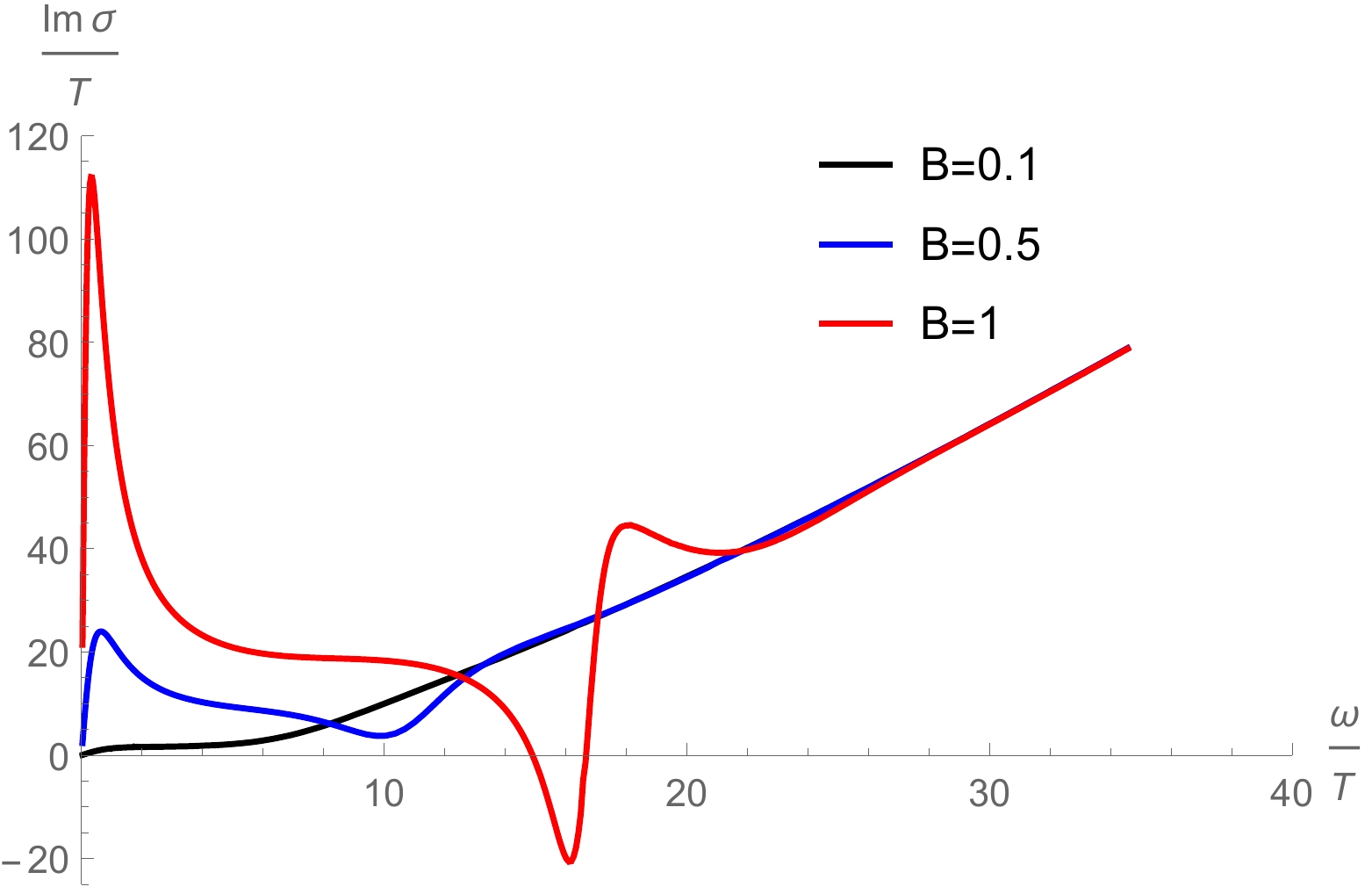} \\
  \end{tabular}
  \caption{\small Real and imaginary parts of the AC longitudinal magnetoconductivity for $M/T=\pi$ and $B/T^2=0.1\pi^2,~0.5\pi^2,~\pi^2$ respectively. }    \label{fig:acM}
\end{figure}

We can solve these equations numerically by integrating from the horizon to the boundary with the boundary condition that the source of $a_t$ and $\phi_2$ is $0$. The conductivity can be calculated from \be \sigma_{zz}=\frac{2v_{z1}}{i\omega v_{z0}}+\frac{i \omega}{2}.\ee Using the fact that the system is invariant under the residual symmetry $a_t \to a_t+i\omega \Lambda,~~ \phi_2 \to \phi_2-q\Lambda \phi$ where $\Lambda$ is a constant independent of $r$, we will be able to generate solutions with $a_t=0$ for each independent numerical solution. Then we can  use the two free parameters at the horizon to generate solutions which has no source of $\phi_2$ at the boundary. In Fig.~\ref{fig:acM} we show the AC longitudinal magnetoconductivity for $M/T=\pi$ and $B/T^2=0.1\pi^2,~0.5\pi^2,~\pi^2$ respectively. We can see from the figures that after adding this axial charge dissipation, the zero frequency pole in the imaginary part indeed vanishes and instead a drude peak develops at small frequency even for $M/T \sim O(1)$, i.e. when the axial charge conservation symmetry is completely broken. As $B$ increases, the height of the drude peak also increases which means that the axial relaxation time increases with $B$. At larger $B$ quasinormal modes start to develop at large values of $\omega$.

As can be seen from the numerics above, with the axial charge dissipation we have a finite DC longitudinal magnetoconductivity. In this case, we can calculate the DC conductivity using the radially conserved quantity \cite{Iqbal:2008by} following \cite{{Donos:2014cya},{Jimenez-Alba:2015awa}}. Consider $\delta V_{\mu}=(0,0,0,0,-E t+v_z(r))$ and $\delta A_{\mu}=(a_t(r),0,0,0,0)$, the equations are now
\be
v_z''+v_z'\bigg(\frac{f'}{f}+\frac{n'}{n}-\frac{h'}{2h}\bigg)+\frac{8\alpha B \sqrt{h}a_t'}{f n}=0
\ee
\be
a_t''+a_t'\bigg(\frac{n'}{n}+\frac{h'}{2h}\bigg)+\frac{8\alpha B v_z'}{n\sqrt{h}}-\frac{2 q^2 a_t \phi^2}{f}=0
\ee
 The radially conserved current is \be J_z(r)=-\frac{f n}{\sqrt{h}}v_z'-8\alpha B a_t,\ee and we have $J_z(\infty)=J_z(r_0)$.
 At the horizon, we have the ingoing boundary condition
 \be v_z(r_0)\simeq -E t-\frac{E}{4\pi T}\ln(r-r_0)\ee and \be a_t\simeq \frac{-4 \alpha B E}{n_0 \sqrt{h_0} q^2 \phi^2(r_0)}.\ee
 Thus we have the DC longitudinal magnetoconductivity \be\label{sigmaanalytic} \sigma_{zz} =J_z(\infty)/E=\frac{n_0}{\sqrt{h_0}}+\frac{32\alpha^2 B^2}{n_0 \sqrt{h_0} q^2 \phi^2(r_0)}.\ee This formula contains two parts of contributions. The first part is ${n_0}/{\sqrt{h_0}}$ which reduces to $\pi T$ in the probe limit. This part now also has a dependence on the background magnetic field. The rest is the second part, which reduces to exact $B^2$ behavior in the probe limit. Due to backreactions, the $B^2$ scaling behavior of this part might also become different.  We numerically checked that the analytic result agrees with our numerical results. With this analytic formula for the DC longitudinal magnetoconductivity, we can reach for arbitrary large $B$ region. Thus we do not need to go to the zero temperature limit to work on the large $B$ behavior and we give the zero temperature background solutions to this system in the appendix. In the following we focus on the small $M$ region where $\tau_5$ is large enough to stay in the hydrodynamic regime.

  \begin{figure}[h]
    \centering
    \begin{tabular}{@{}cccc@{}}
    \includegraphics[width=.47\textwidth]{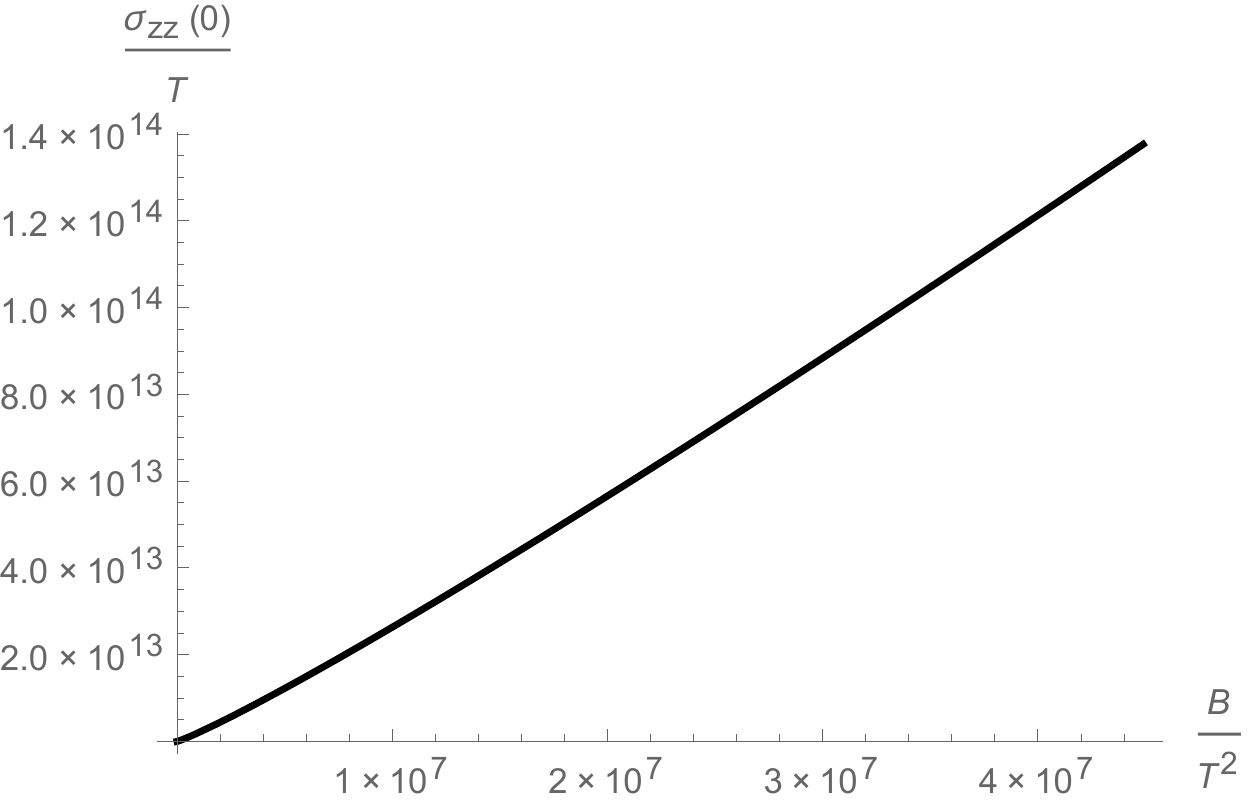}&  \includegraphics[width=.47\textwidth]{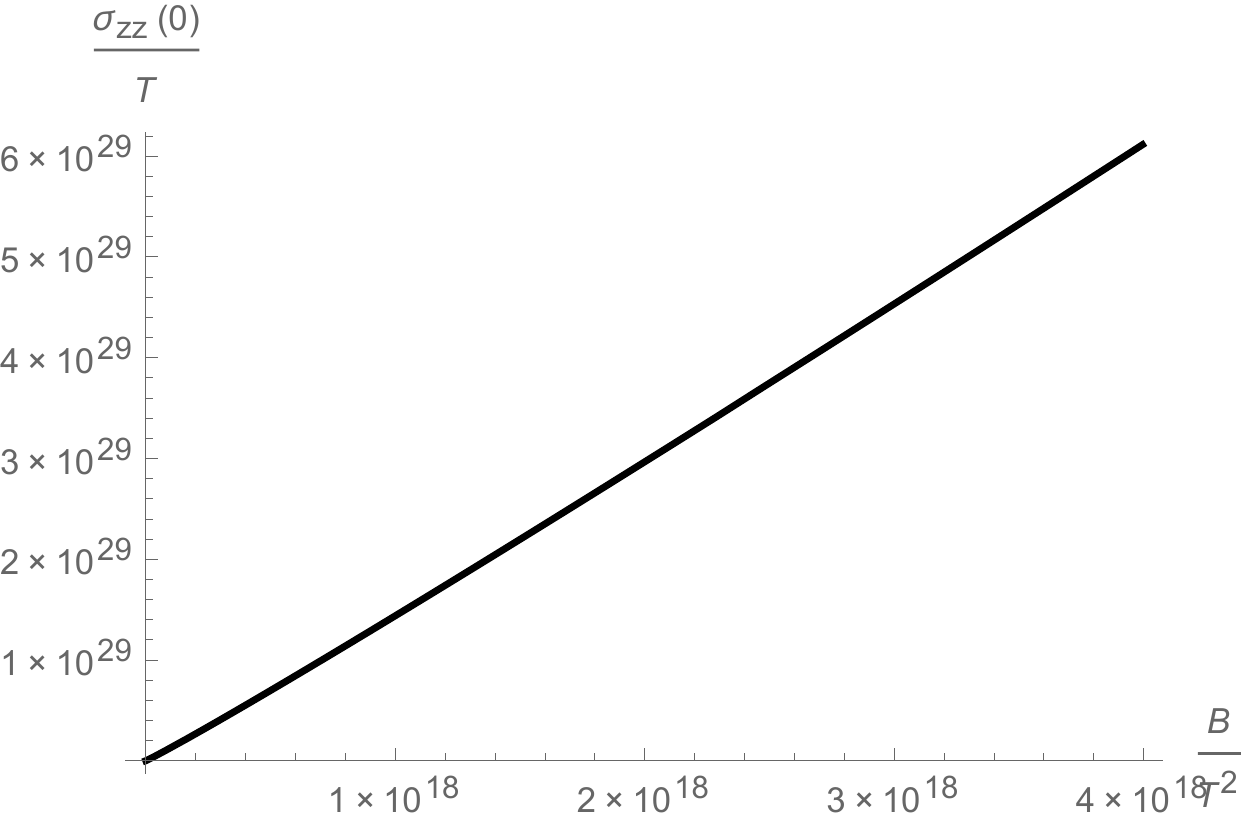} \\
\includegraphics[width=.47\textwidth]{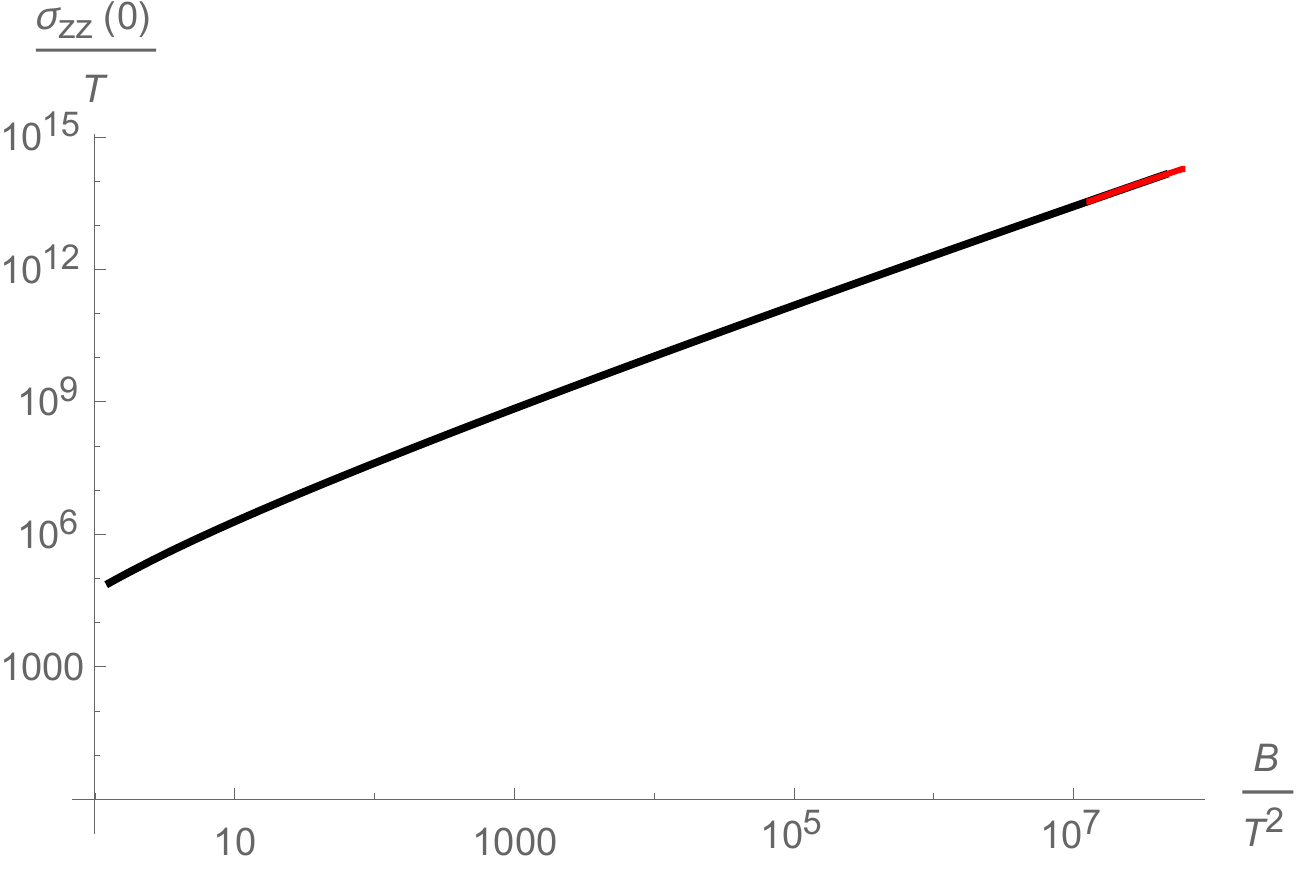}&  \includegraphics[width=.47\textwidth]{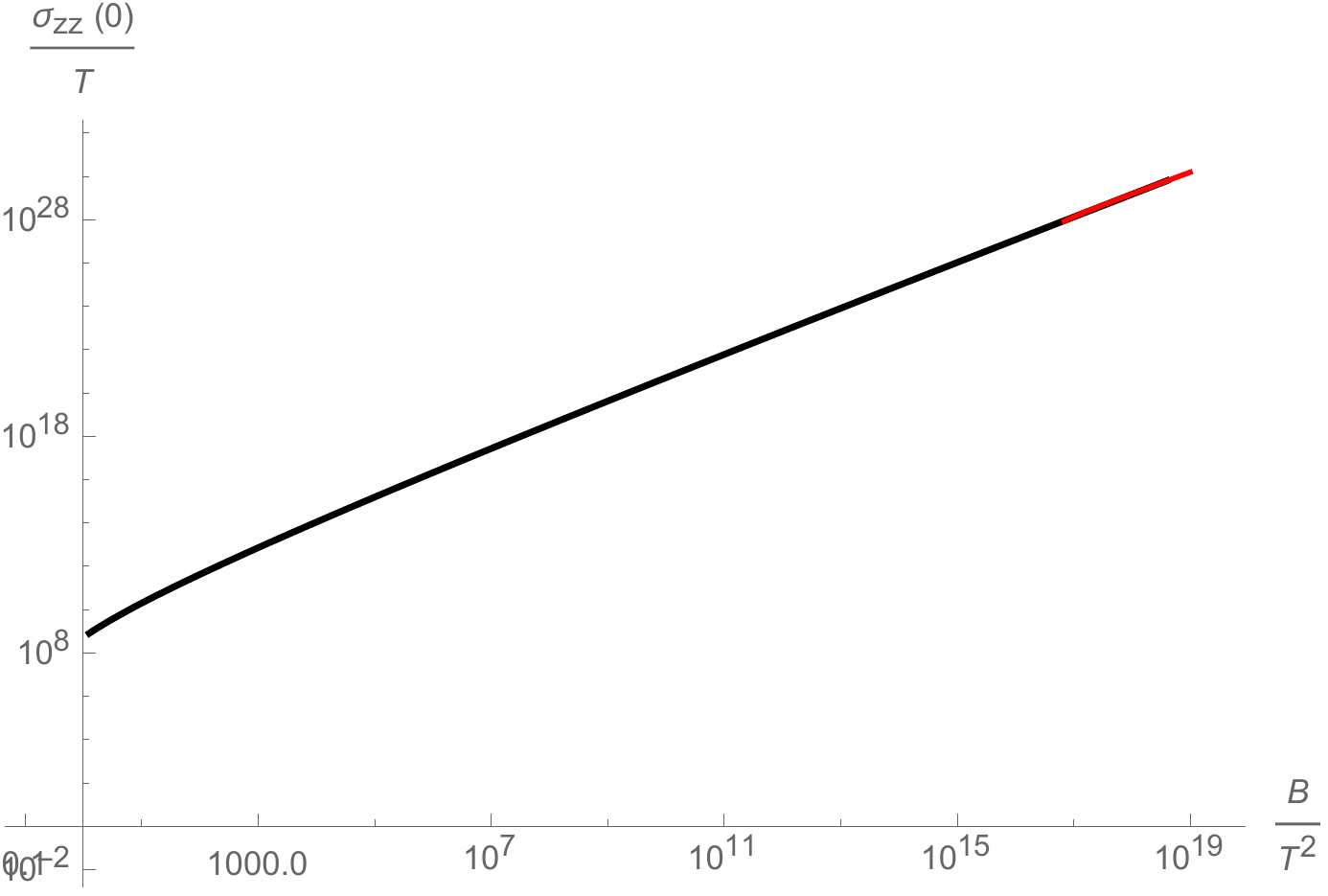} \\
  \end{tabular}
  \caption{\small Top: the DC longitudinal magnetoconductivity $\sigma_{zz}(0)$ as a function of $B/T^2$ for fixed $M/T=0.005\pi$ (left) and $M/T=0.00005\pi$ (right). Bottom: log-log plots for the same figures above. The red lines are slope $1$ functions at the large $B$ region, which indicates that at large $B/T^2$, $\sigma_{zz}(0)$ is indeed a linear function of $B/T^2$ .}    \label{fig:simgazzfixingM}
\end{figure}

Different from the universal $B^2$ behavior of the probe limit, after taking into account the backreactions to the geometry, $n_0$, $h_0$ and $\phi_{0}$ now all depend on both $B$ and $M$. At small $B/T^2$, the leading order dependence on $B$ in all these parameters should be the same as the probe limit and deviations from the probe limit only arise at larger $B/T^2$ and $\lambda$. In the following we mainly focus on the large $B/T^2$ behavior of the DC longitudinal magnetoconductivity at fixed small values of $M/T$ and large $\lambda$. In Fig.~\ref{fig:simgazzfixingM} we plot the DC longitudinal magnetoconductivity $\sigma_{zz}(0)$ as a function of $B/T^2$ at fixed $M/T=0.005\pi$ (left) and $M/T=0.00005\pi$ (right) at $\lambda=200$. At large $B/T^2$, $\sigma_{zz}(0)$ grows linearly in $B/T^2$. To analyze the scaling of $\sigma_{zz}(0)$ on $B$ more explicitly, it is better to study the two parts in the analytic formula (\ref{sigmaanalytic}) separately.

 \begin{figure}[h]
    \centering
    \begin{tabular}{@{}cccc@{}}
    \includegraphics[width=.47\textwidth]{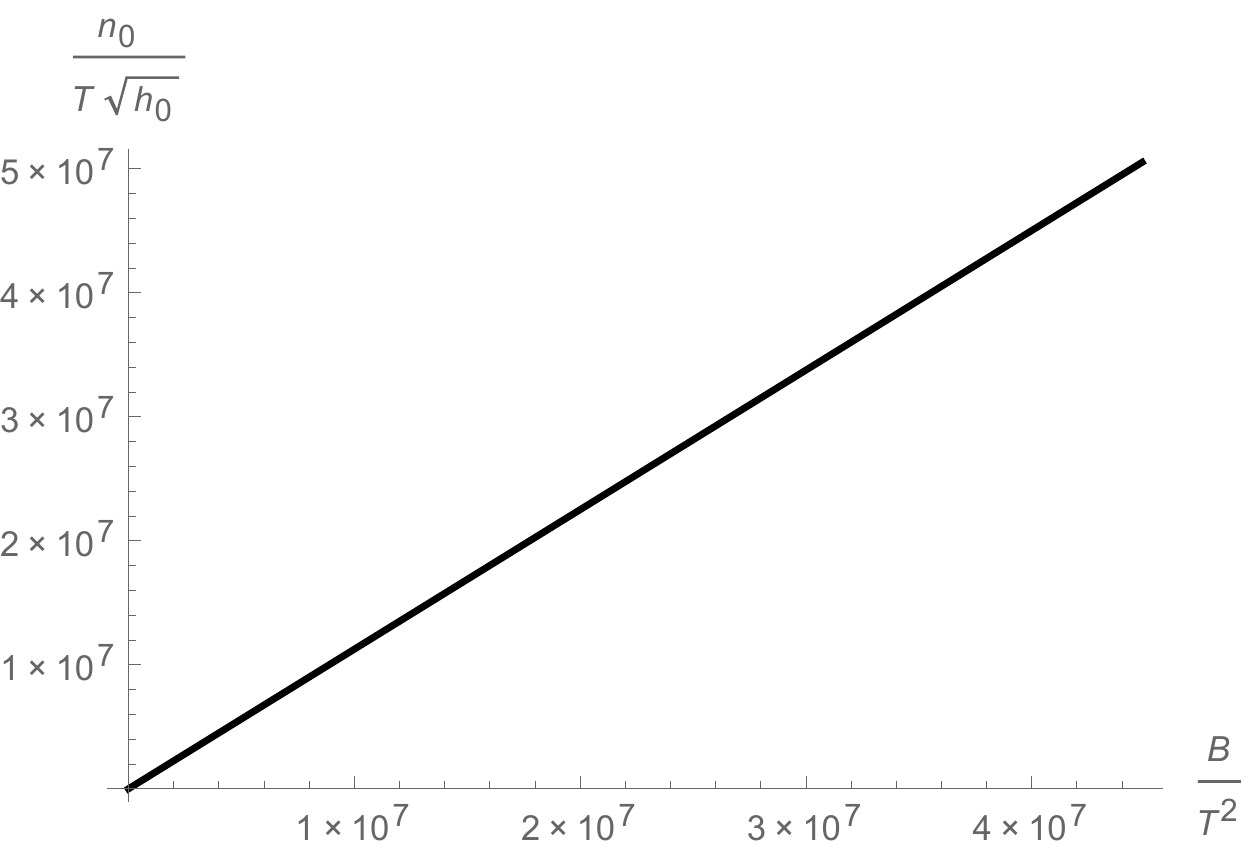} &\includegraphics[width=.47\textwidth]{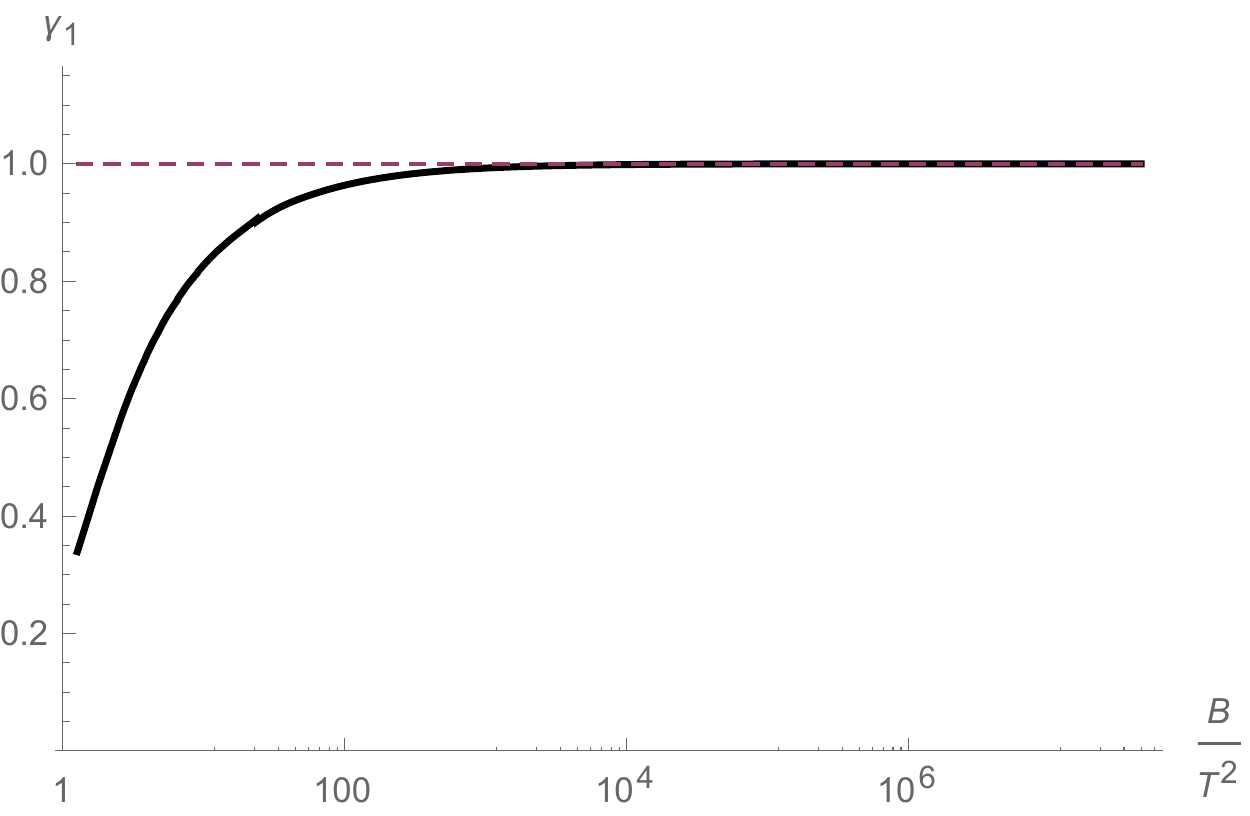}\\
  \includegraphics[width=.47\textwidth]{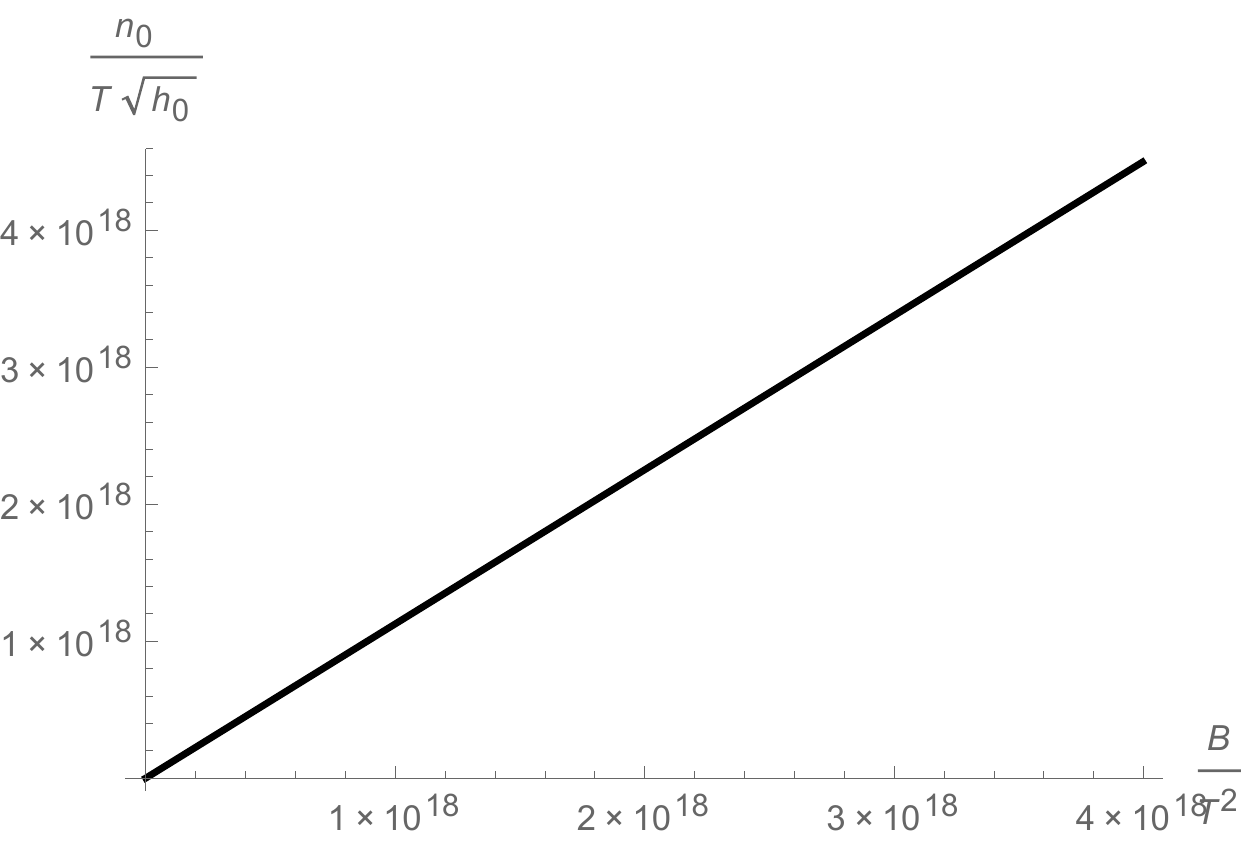}&  \includegraphics[width=.47\textwidth]{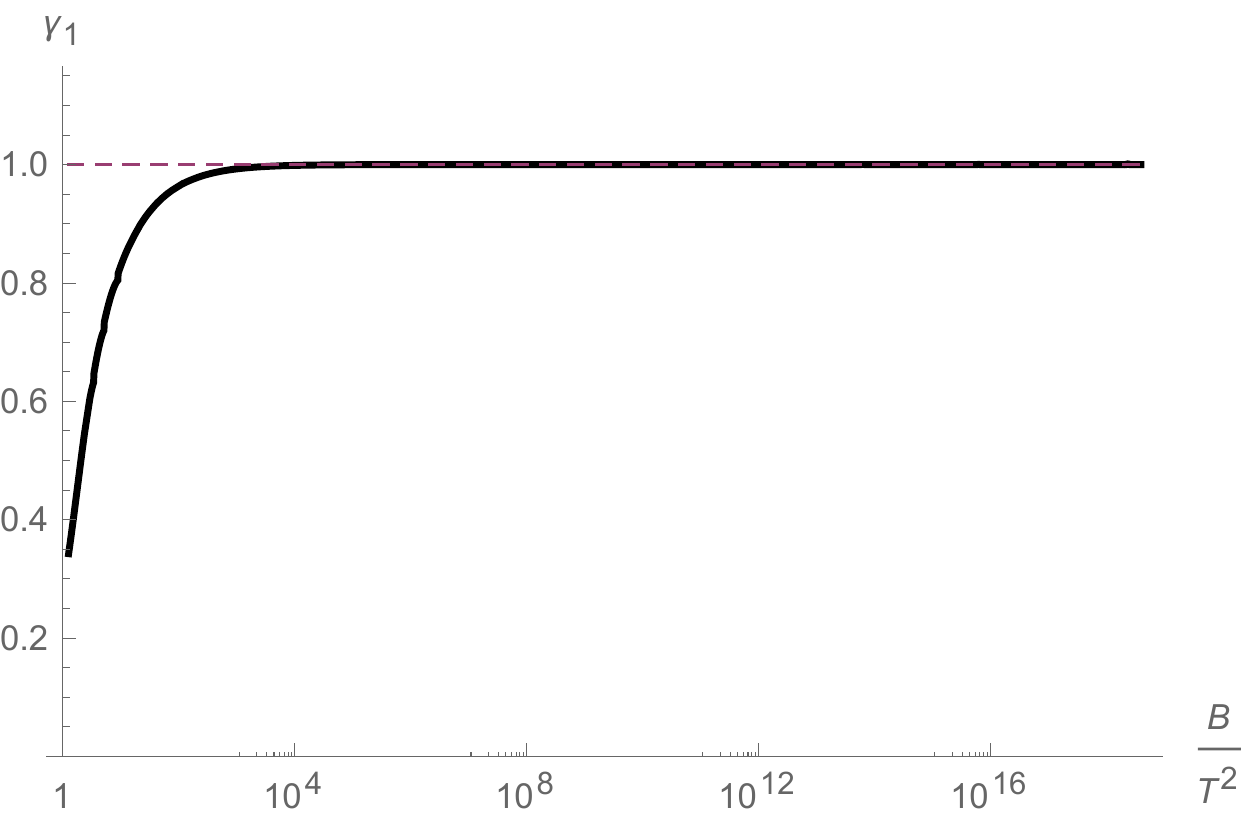} \\
  \end{tabular}
  \caption{\small The dependence of the first part ${n_0}/{\sqrt{h_0}}$ in the analytic formula for the DC longitudinal magnetoconductivity (\ref{sigmaanalytic}) as well as $\gamma_1=B({n_0}/{\sqrt{h_0}})'/({n_0}/{\sqrt{h_0}})$ as a function of $B/T^2$ at fixed $M/T=0.005\pi$ (top) and $M/T=0.00005\pi$ (bottom) at $\lambda=200$.}    \label{fig:simga1fixingM}
\end{figure}

We denote the scaling exponent of $B/T^2$ in the first part ${n_0}/{\sqrt{h_0}}$ in the formula (\ref{sigmaanalytic}) as $\gamma_f$, i.e. ${n_0}/{\sqrt{h_0}}\simeq c(M/T) (B/T^2)^{\gamma_f}$ at large $B/T^2$. In numerics we can get the value of the scaling exponent $\gamma_f$ using $\gamma_1=B({n_0}/{\sqrt{h_0}})'/({n_0}/{\sqrt{h_0}})$ and by definition this value that we obtained only has the meaning of the scaling exponent when it remains a constant in a finite region of $B/T^2$. In Fig.~\ref{fig:simga1fixingM} we show the dependence of ${n_0}/{\sqrt{h_0}}$ and the value of $\gamma_1$ at fixed $M/T=0.005\pi$ (top figure) and $M/T=0.00005\pi$ (bottom) separately. Due to numerical constraints, we can reach for much larger values of $B/T^2$ for the $M/T=0.00005\pi$ case. As we can see from the figure, the value of $\gamma_1$ reaches a constant $1$ at large $B$, indicating a scaling behavior at large $B$ with scaling exponent being $\gamma_f=1$, in contrast to the behavior of ${n_0}/{\sqrt{h_0}}\simeq \pi T+ c_{s} B^2$ at small $B/T^2$ and $\sqrt{\lambda} B/T^2$, where $c_s$ denotes a constant independent of $B$. However, at large $B/T^2$ this term is not the main contribution to $\sigma_{zz}(0)$ in the formula (\ref{sigmaanalytic}) as this term is much smaller than the second part.

  \begin{figure}[h]
    \centering
    \begin{tabular}{@{}cccc@{}}
    \includegraphics[width=.47\textwidth]{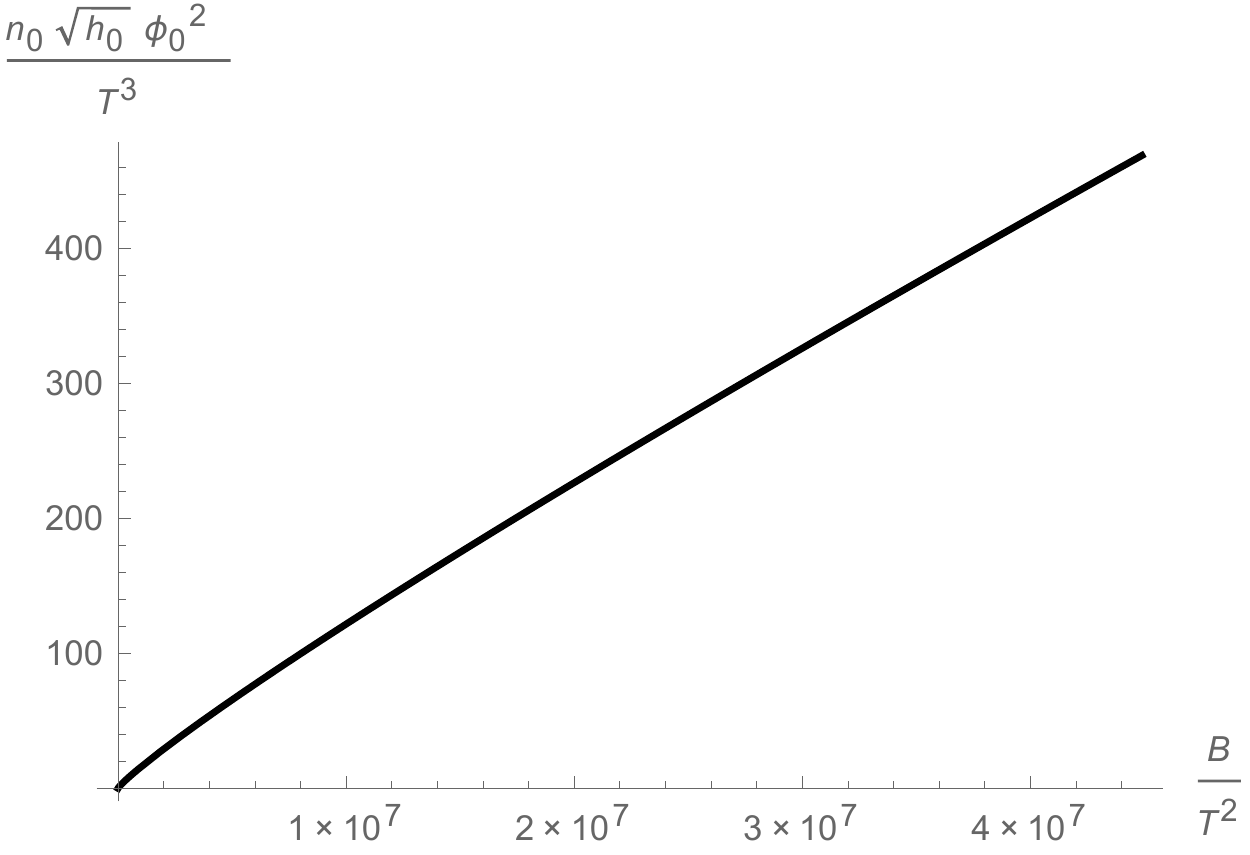} &\includegraphics[width=.47\textwidth]{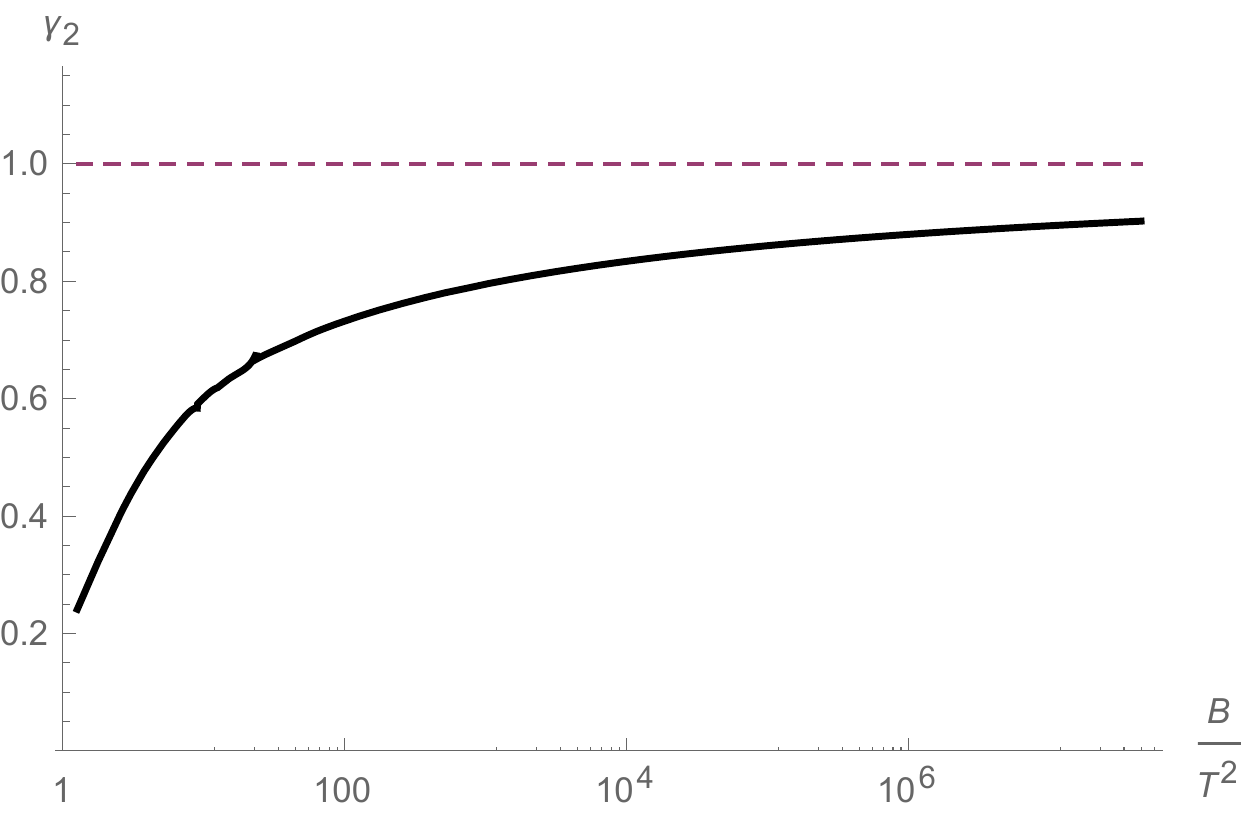}\\
  \includegraphics[width=.47\textwidth]{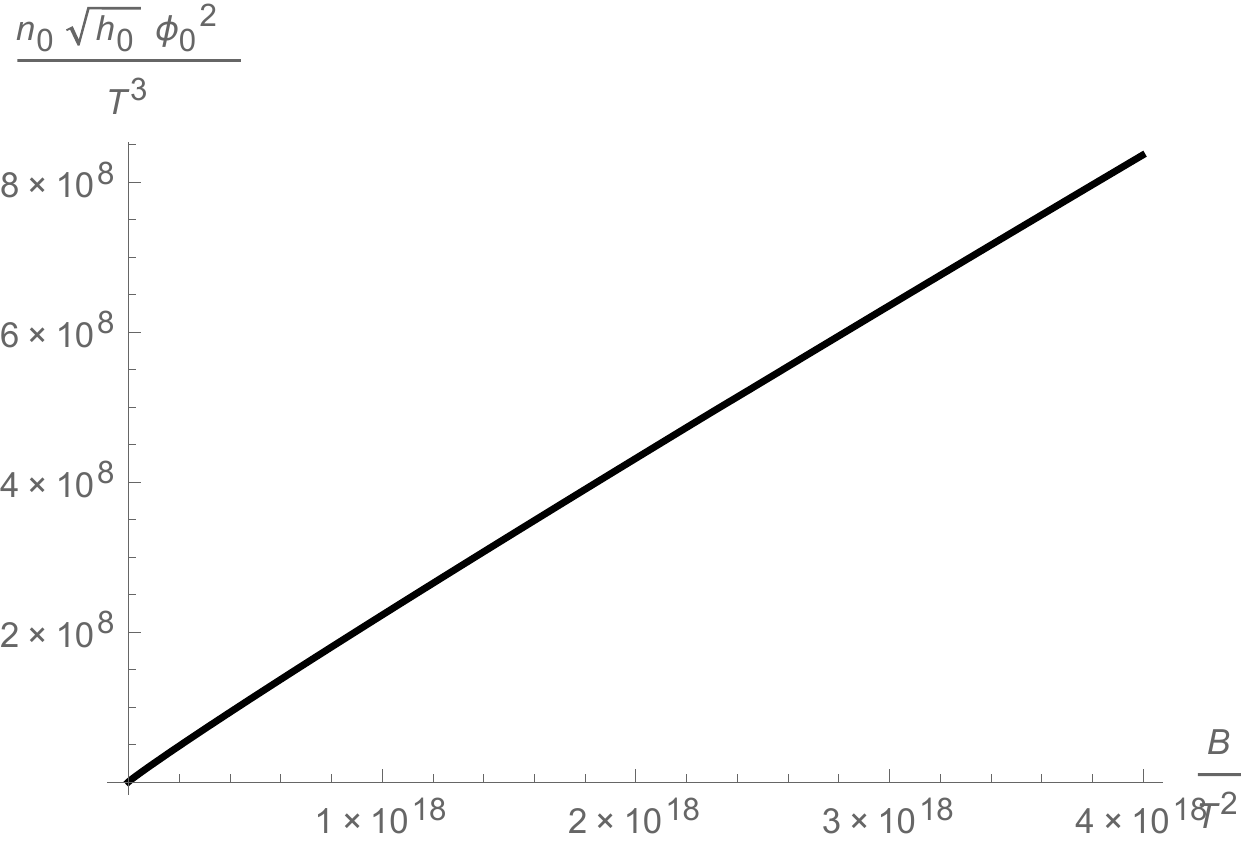}&  \includegraphics[width=.47\textwidth]{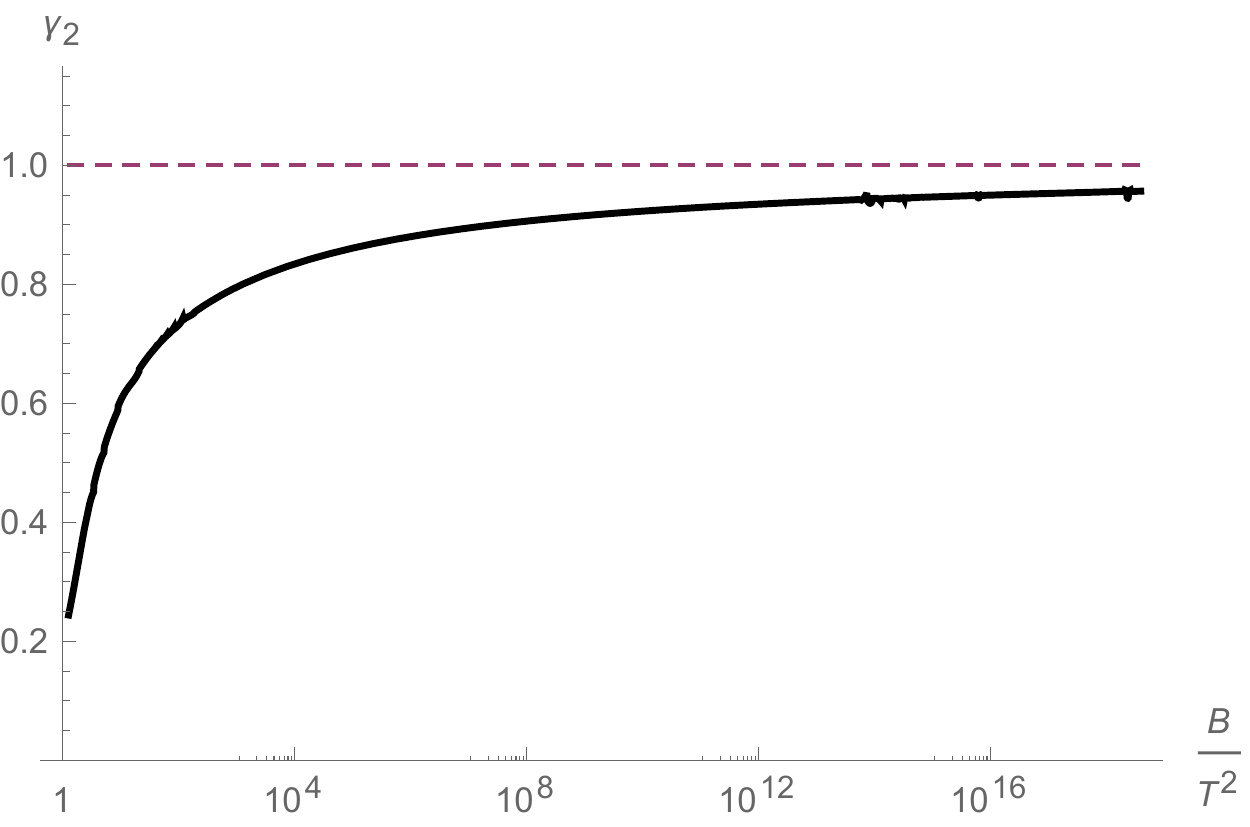} \\
  \end{tabular}
  \caption{\small The dependence of the denominator $\frac{32\alpha^2 B^2}{n_0 \sqrt{h_0} q^2 \phi^2(r_0)}$ in the second part of the analytic formula for the DC longitudinal magnetoconductivity (\ref{sigmaanalytic}) as well as $\gamma_2=B(n_0 \sqrt{h_0}\phi^2(r_0))'/(n_0 \sqrt{h_0}\phi^2(r_0))$ as a function of $B/T^2$ at fixed $M/T=0.005\pi$ (top) and $M/T=0.00005\pi$ (bottom) at $\lambda=200$.}    \label{fig:simga2fixingM}
\end{figure}

In the second part of formula (\ref{sigmaanalytic}), the numerator in $\frac{32\alpha^2 B^2}{n_0 \sqrt{h_0} q^2 \phi^2(r_0)}$ is exactly $B^2$ and the full dependence of this term on $B$ is determined by the dependence on $B$ in the denominator $n_0 \sqrt{h_0}\phi^2(r_0)$. We denote the scaling exponent of $B$ in  $n_0 \sqrt{h_0}\phi^2(r_0)$ as $\gamma_s$ for large $B$, i.e.  $n_0 \sqrt{h_0}\phi^2(r_0)\simeq c(M,T) B^{\gamma_s}$. In numerics we can get the value of the scaling exponent $\gamma_s$ using $\gamma_2=B(n_0 \sqrt{h_0}\phi^2(r_0))'/(n_0 \sqrt{h_0}\phi^2(r_0))$ and by definition this value that we obtained only has the meaning of the scaling exponent when it remains a constant in a finite region of $B/T^2$. In Fig.~\ref{fig:simga2fixingM} we plot the dependence on B of $n_0 \sqrt{h_0}\phi^2(r_0)$ as well as $\gamma_2$ at fixed $M/T=0.005\pi$ (top figure) and $M/T=0.00005\pi$ (bottom) separately. For the case of $M/T=0.005\pi$, due to numerical constraints, we cannot reach very large $B/T^2$ region, but we can already see that $\gamma_2$ is approaching $1$ slowly as $B$ becomes larger, indicating a scaling behavior with $\gamma_s=1$. In the figure of $M/T=0.00005\pi$ we can already see that $\gamma_2$ almost goes to $1$ at large $B/T^2$\footnote{However we cannot at present rule out a power law which deviates slightly from $1$ due to the numerical constraints.}. Substituting this scaling behavior into the second part of the analytic formula for $\sigma_{zz}(0)$ (\ref{sigmaanalytic}), we can see that the second part in the formula also goes linearly in $B$ at large $B/T^2$, compared to the $B^2$ behavior of the small $B/T^2$ limit. Note that the second part is much larger than the first part in the analytic formula of the DC conductivity. Thus we can see that after considering backreaction effects, the DC longitudinal magnetoconductivity is linear in $B$ at large $B/T^2$, which is different from the exact $B^2$ behavior of the probe limit. This scaling behavior coincides with the weakly coupled kinetic result qualitatively \cite{Nielsen:1983rb,{sonspivak}}. In one of the experiments \cite{kimprl}, the same scaling behavior was also found.


 From the hydrodynamic formula, at small $B/T^2$ and large axial charge relaxation time, the DC longitudinal magnetoconductivity obeys the following formula
 \be\label{hydroformula}
 \sigma_{DC}=\sigma_E+(8\alpha B)^2\frac{\tau_5}{\chi_5},
 \ee
 where $\tau_5$ is the axial charge relaxation time. We can calculate $\tau_5$ and $\chi_5$ numerically using this same setup but with different boundary conditions at the asymptotic $AdS_5$ boundary. For $\chi_5$, we choose the boundary condition that $v_z$ and $\phi_2$ are sourceless at the boundary. We can also simplify the three equations for perturbations into one equation of motion for $a_t$ at zero frequency \be a_t''+a_t'\Big(\frac{n'}{n}+\frac{h'}{2 h}\Big)-\frac{64\alpha^2 B^2 a_t}{n^2f}-\frac{2 q^2 \phi^2 a_t}{f}=0.\ee

  \begin{figure}[h]
    \centering
    \begin{tabular}{@{}cccc@{}}
    \includegraphics[width=.47\textwidth]{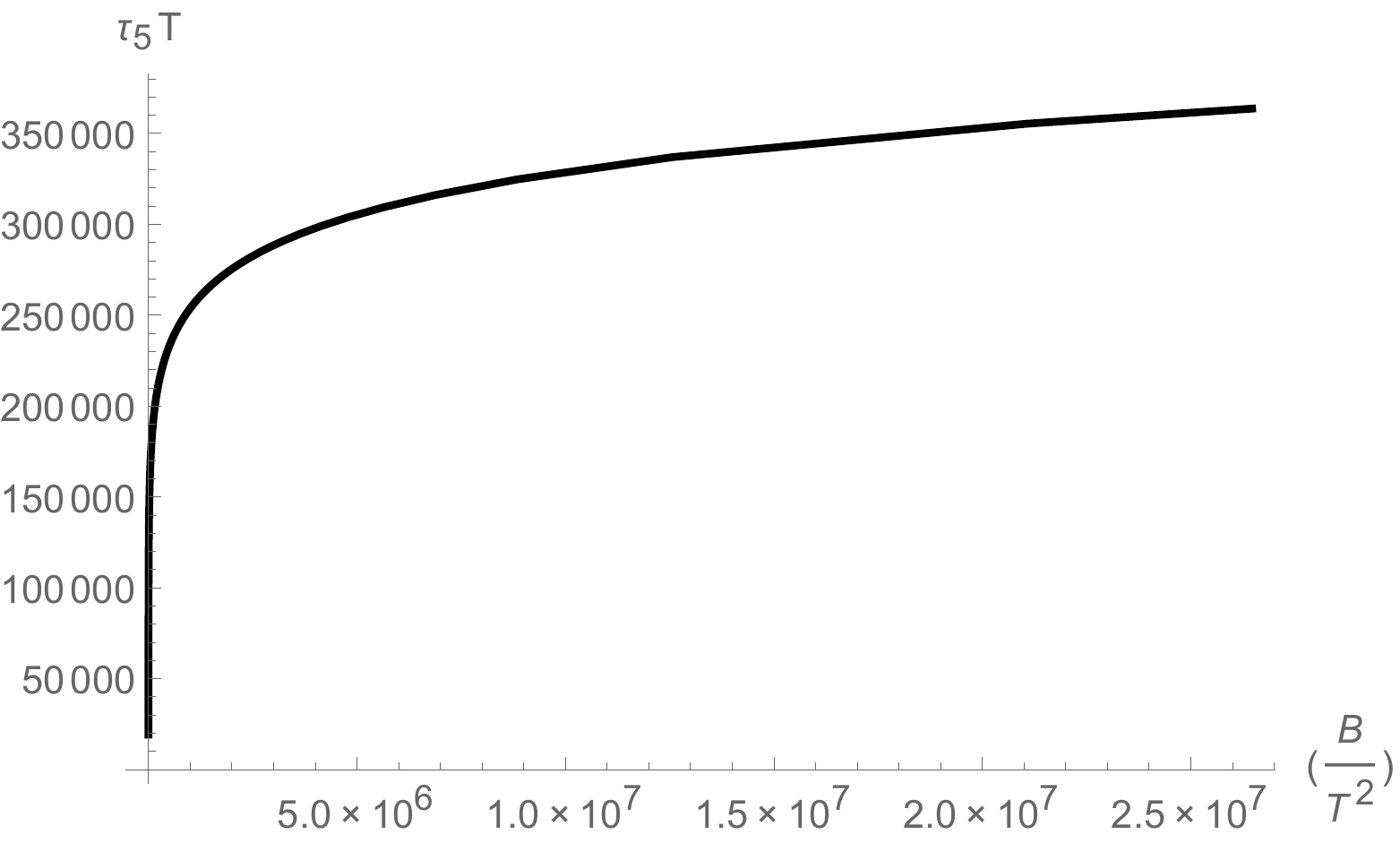} &\includegraphics[width=.47\textwidth]{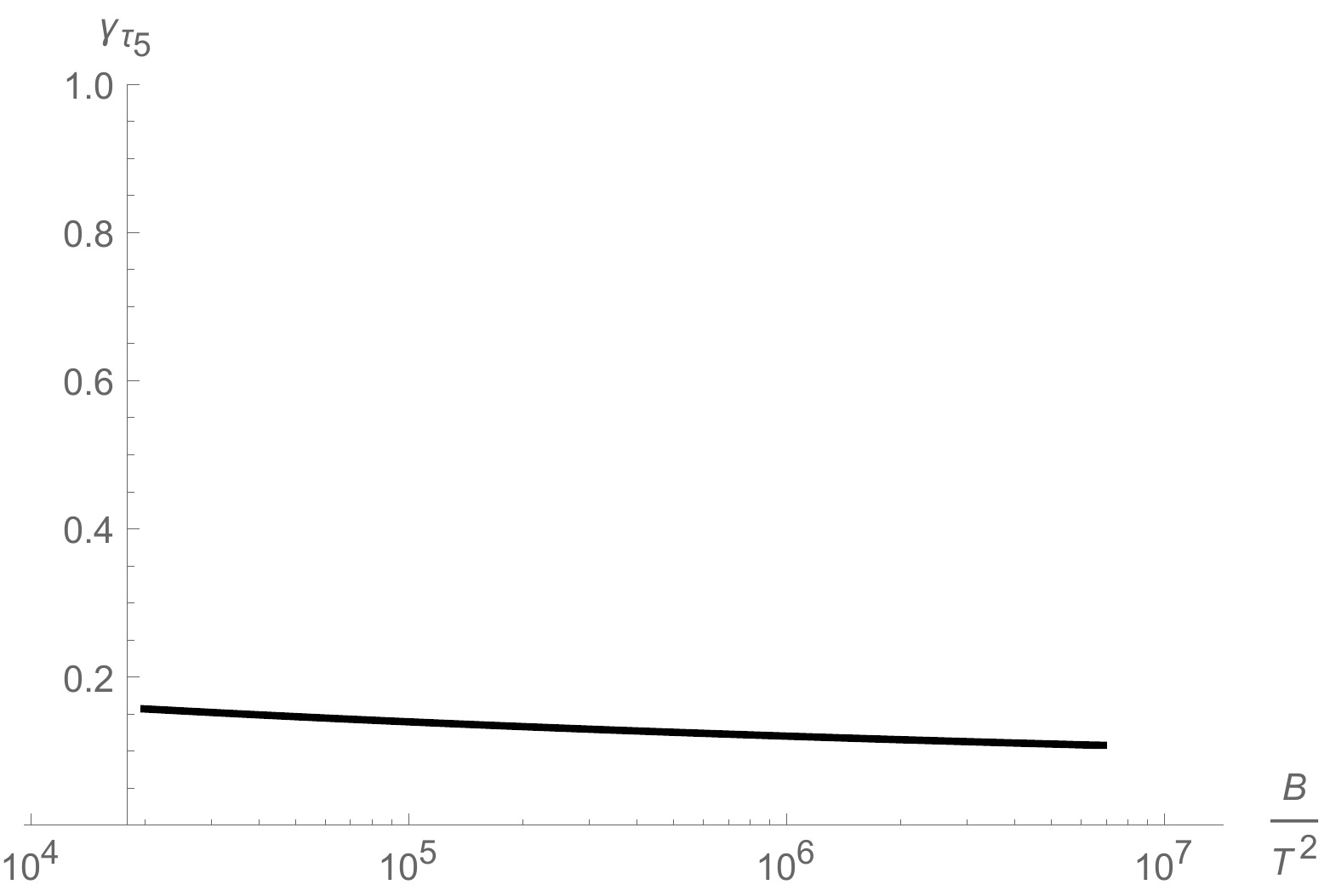}\\
  \includegraphics[width=.47\textwidth]{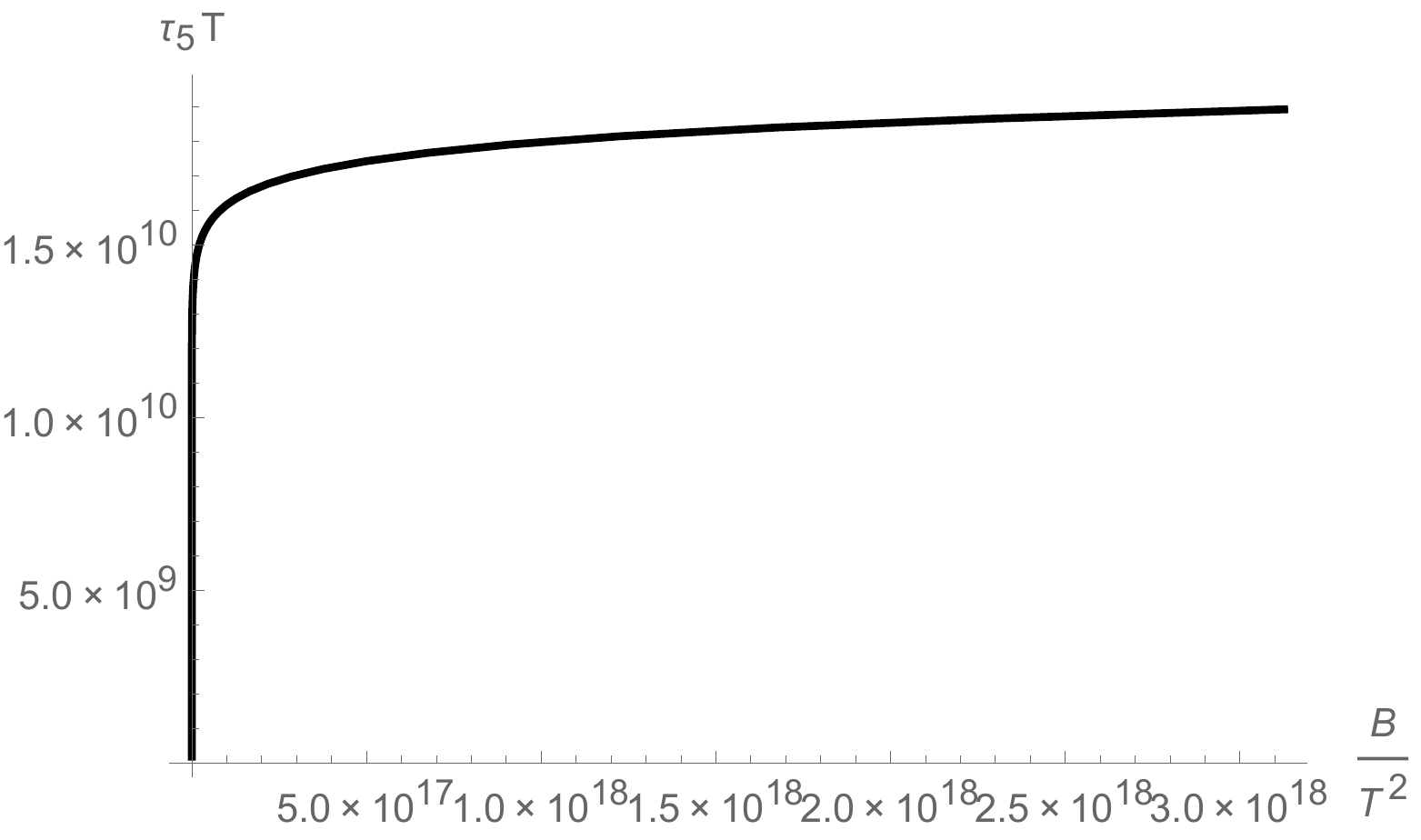}&  \includegraphics[width=.47\textwidth]{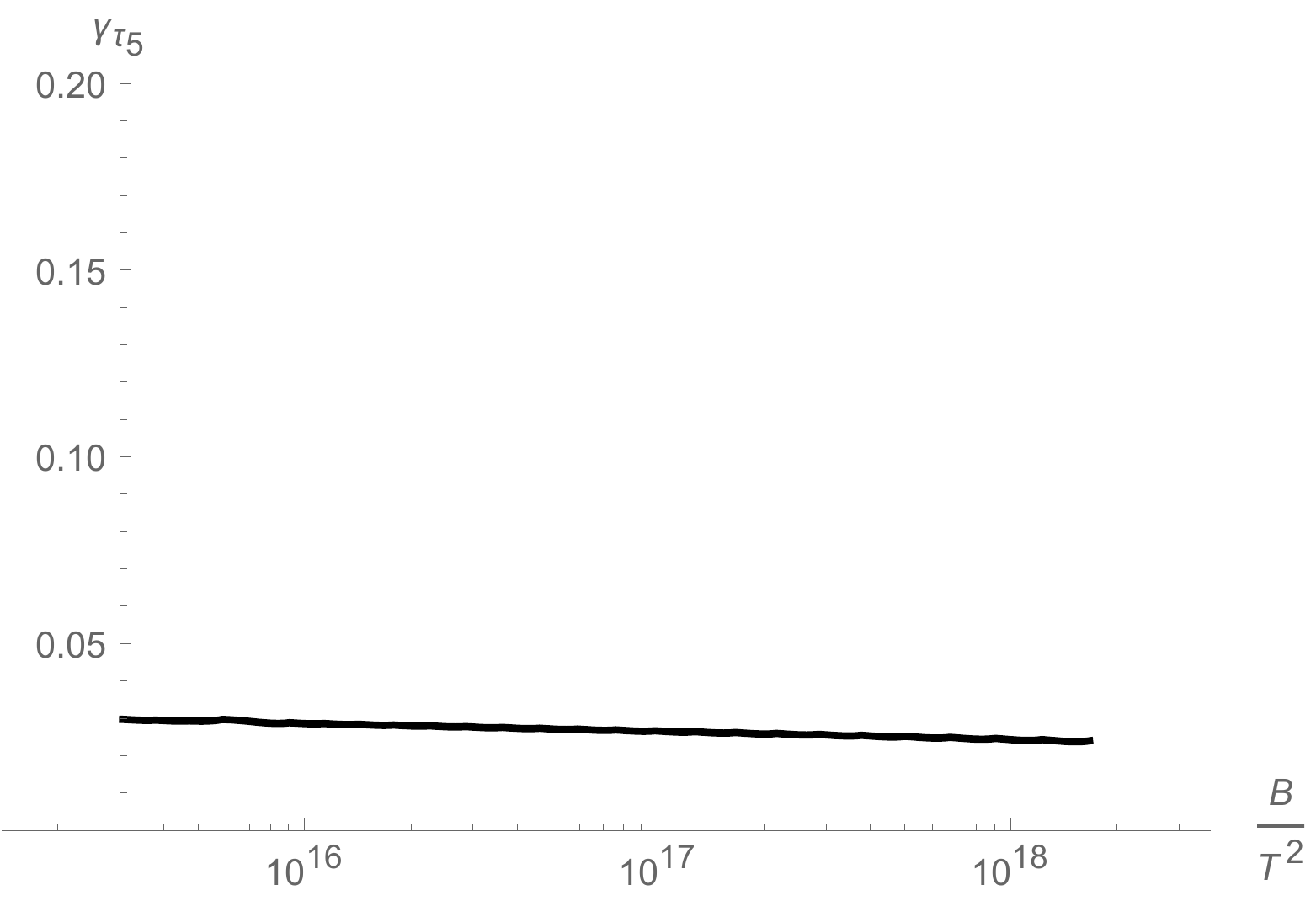} \\
  \end{tabular}
  \caption{\small The dependence of $\tau_5$ and its scaling exponent $\gamma_{\tau_5}$ at large $B$ on $B/T^2$ for two fixed values of $M/T=0.005\pi\text{ (top)},~0.00005\pi$ (bottom), $\lambda=200$ and $\alpha=1$.}    \label{fig:tau5}
\end{figure}

$\tau_5$ can be determined from the zero momentum quasinormal mode under the boundary condition that all three fields $v_z$, $a_t$ and $\phi_2$ are sourceless at the boundary. When we find a pure imaginary quasinormal mode at frequency $-I \omega_I$ we can get $\tau_5=1/\omega_I$. The detailed procedure of this calculation can be found in \cite{Jimenez-Alba:2015awa}. Here we show the numerical results for these two quantities. In Fig.~\ref{fig:tau5}, we show the dependence of $\tau_5$ and its scaling exponent $\gamma_{\tau_5}$ at large $B$ ($\tau_5\simeq c_{\tau_5} B^{\gamma_{\tau_5}}$) on $B/T^2$ for two fixed values of $M/T=0.005\pi\text{ (top)},~0.00005\pi$ (bottom), $\lambda=200$ and $\alpha=1$. Note that in the figure, we defined $\gamma_{\tau_5}=B \tau_5'/\tau_5$, which only has the meaning of the scaling exponent when it reaches a constant in a certain region of $B$. We can see that $\tau_5$ increases as $B$ increases and reaches a finite and constant value at large $B/T^2\to\infty$, i.e. at $B/T^2\to \infty$, the scaling exponent $\gamma_{\tau_5}\to 0$\footnote{We cannot rule out the possibility that the scaling exponent is slightly above $0$ due to the numerical constraints.}. This means that at fixed $M$, there will be an upper limit in $\tau_5$ no matter how large the magnetic field is and this is very different from the probe limit result, where at large $B/T^2$, $\gamma_{\tau_5}\to 1$ and $\tau_5$ diverges at infinite $B/T^2$. We will see later that this caused the deviation in the dependence of the DC longitudinal magnetoconductivity on $B$ at $B/T^2\to\infty$ from the probe limit result.

 \begin{figure}[h]
    \centering
    \includegraphics[width=.57\textwidth]{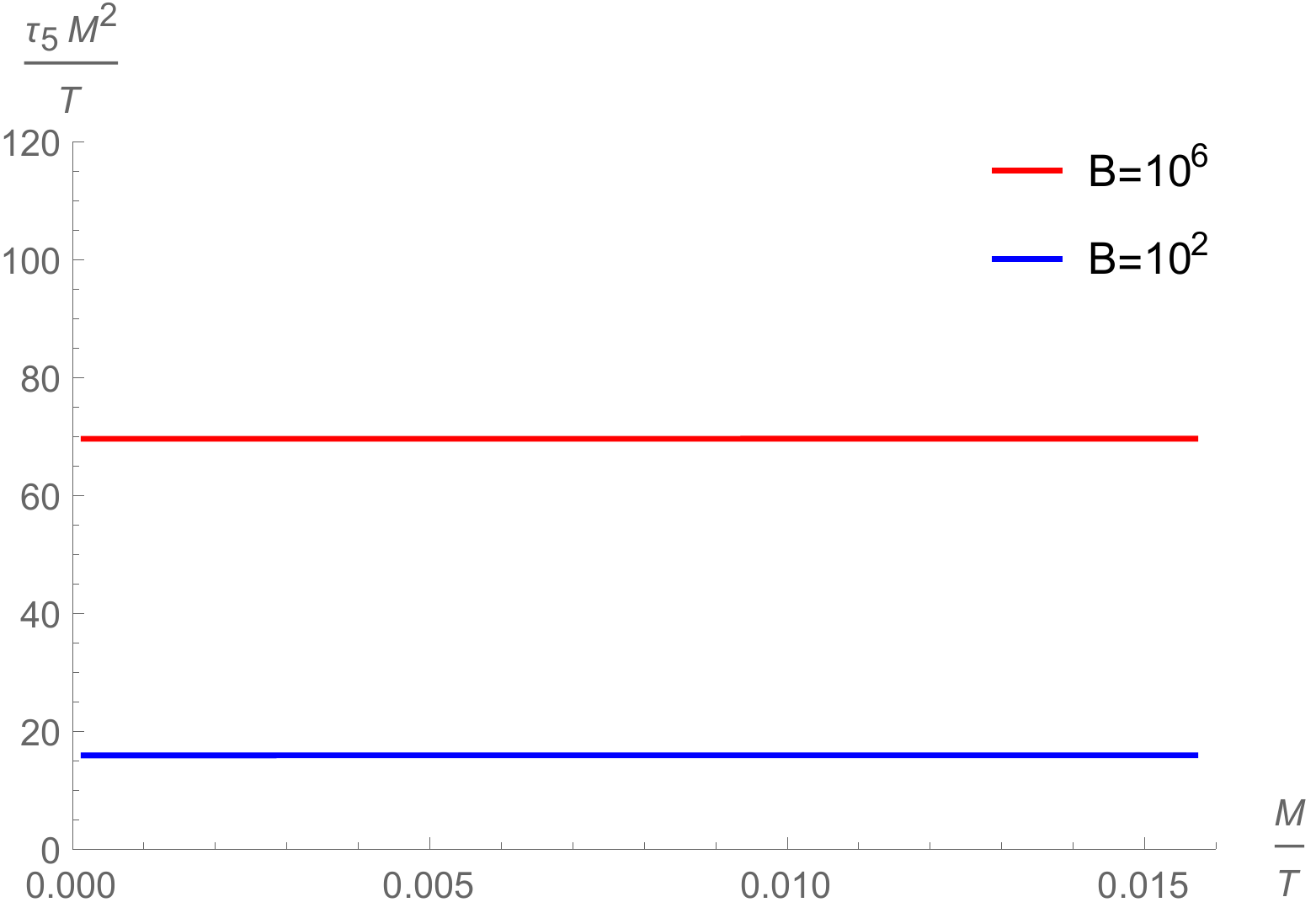}
  \caption{\small For two large and fixed values of $B/T^2= 25\pi^2,~2.5\times 10^5 \pi^2$, $\tau_5 M^{2}$ is a constant at small $M/T$. Here $\lambda=200$ and $\alpha=1$.}    \label{tauMrelation}
\end{figure}

 At small $B$ it is expected that $\tau_5\sim M^{-2}$ at small $M/T$, which is the result from the probe limit. Here we show in Fig.~\ref{tauMrelation} that at two large and fixed values of $B/T^2$, we still have $\tau_5\sim M^{-2}$ at small $M/T$. The holographic axial charge relaxation time and its property was also studied recently in a top down model in \cite{Guo:2016nnq} in AdS/QCD.

  \begin{figure}[h]
    \centering
    \begin{tabular}{@{}cccc@{}}
    \includegraphics[width=.47\textwidth]{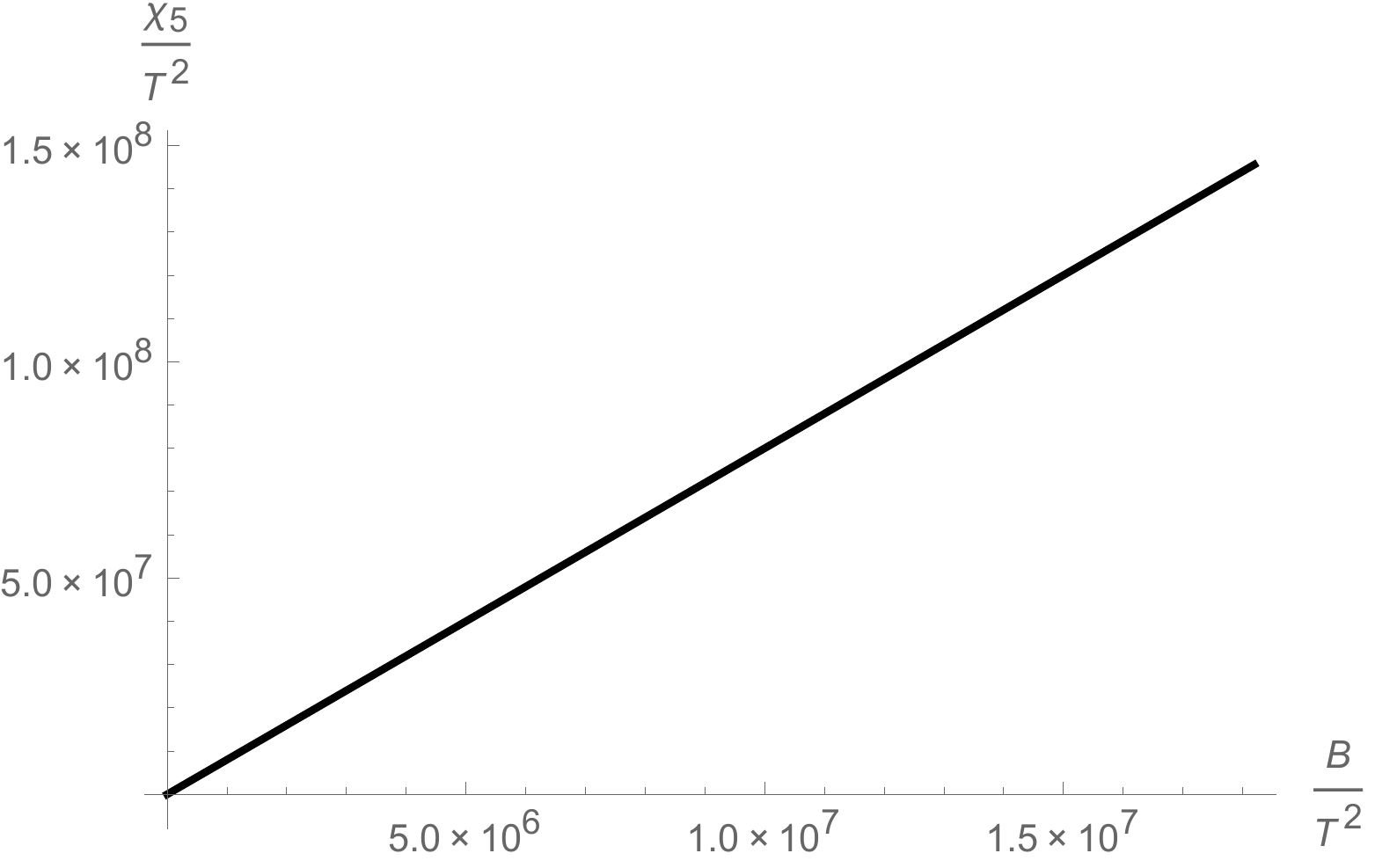} &\includegraphics[width=.47\textwidth]{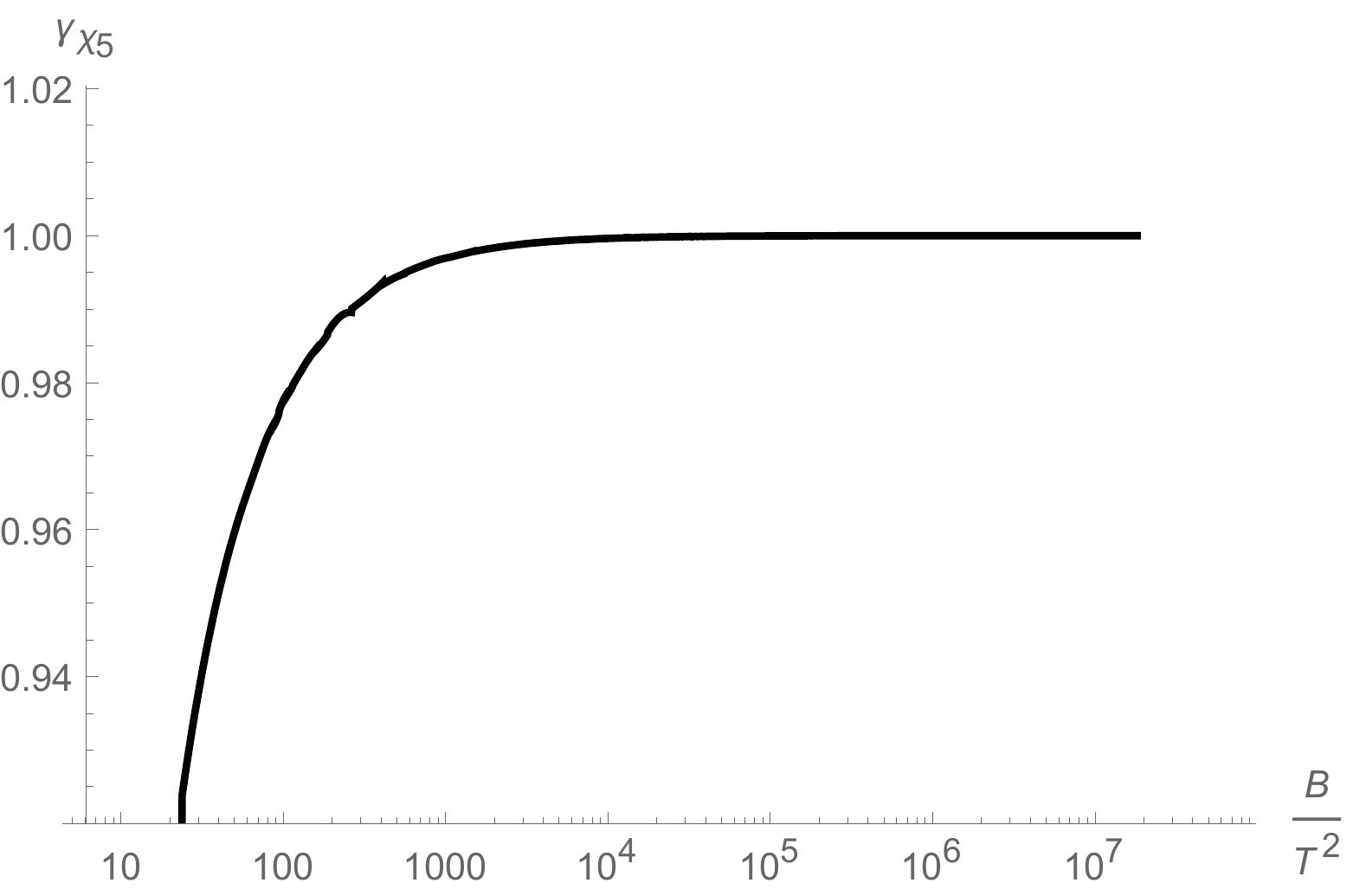}\\
  \includegraphics[width=.47\textwidth]{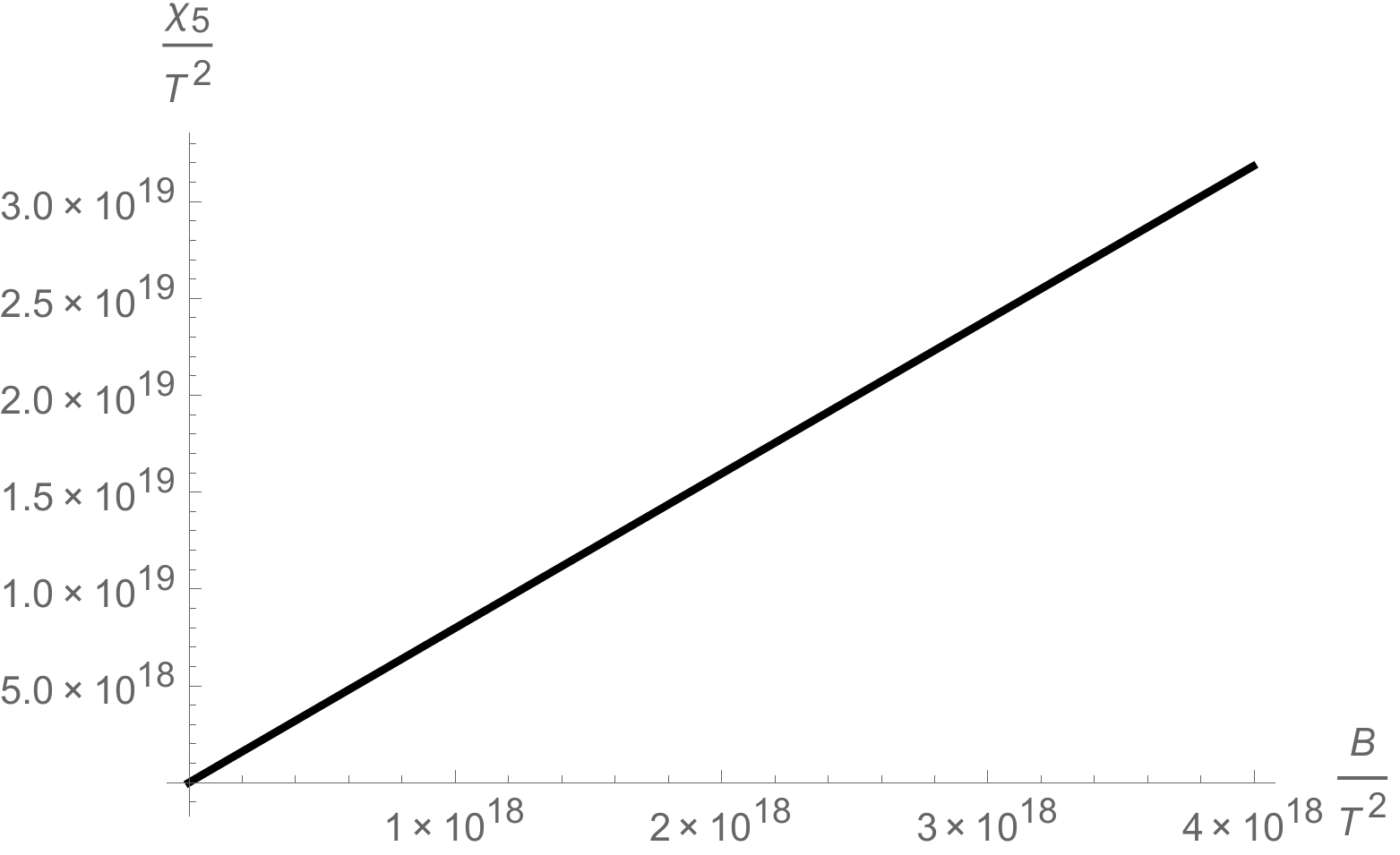}&  \includegraphics[width=.47\textwidth]{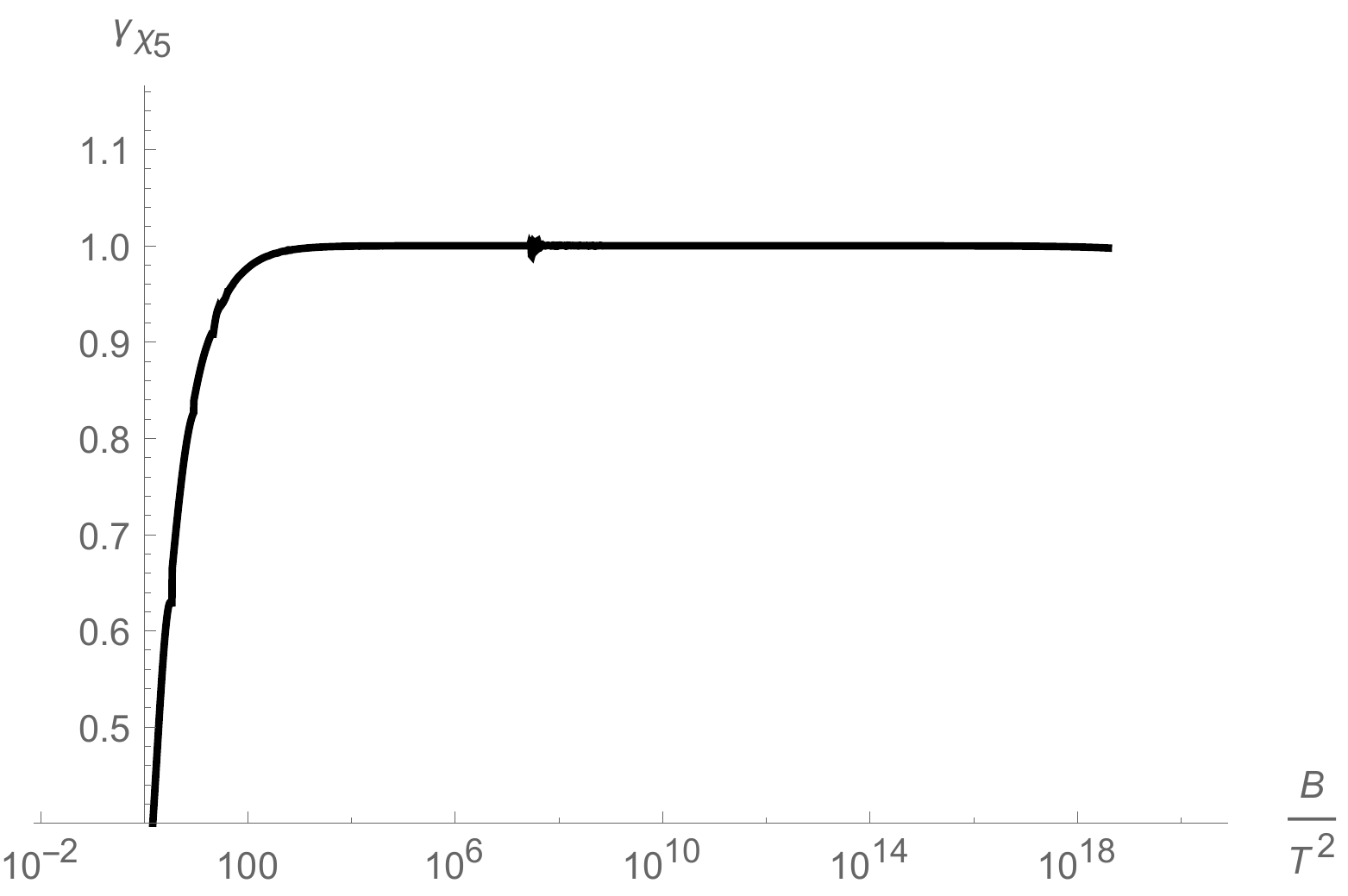} \\
  \end{tabular}
  \caption{\small The dependence of $\chi_5$ and its scaling exponent $\gamma_{\chi_5}$ at large $B$ on $B/T^2$ for two fixed values of $M/T=0.005\pi\text{ (top)},~0.00005\pi$ (bottom), $\lambda=200$ and $\alpha=1$.}    \label{fig:chi5}
\end{figure}

\begin{figure}[h]
    \centering
    \begin{tabular}{@{}cccc@{}}
    \includegraphics[width=.47\textwidth]{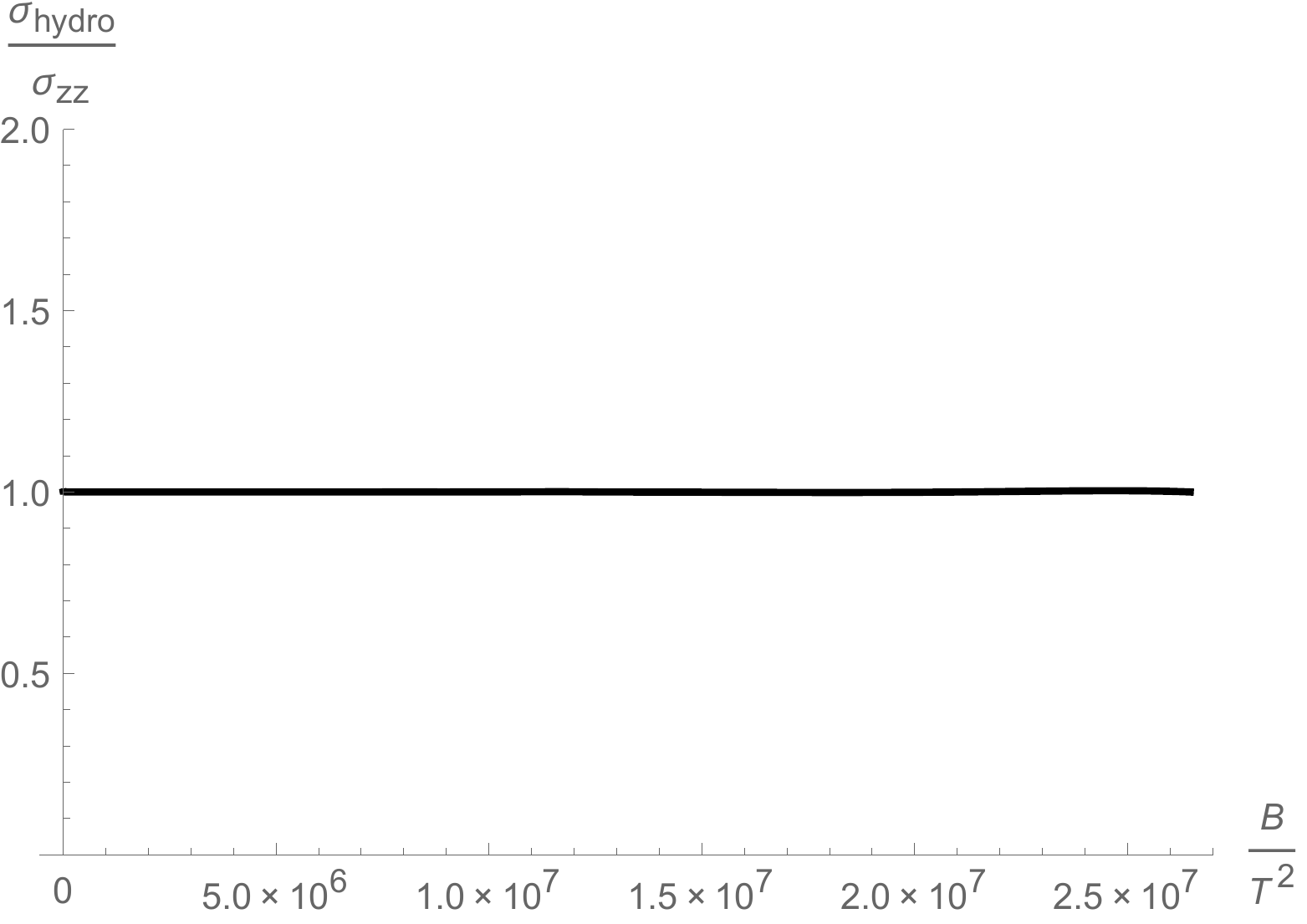} &\includegraphics[width=.47\textwidth]{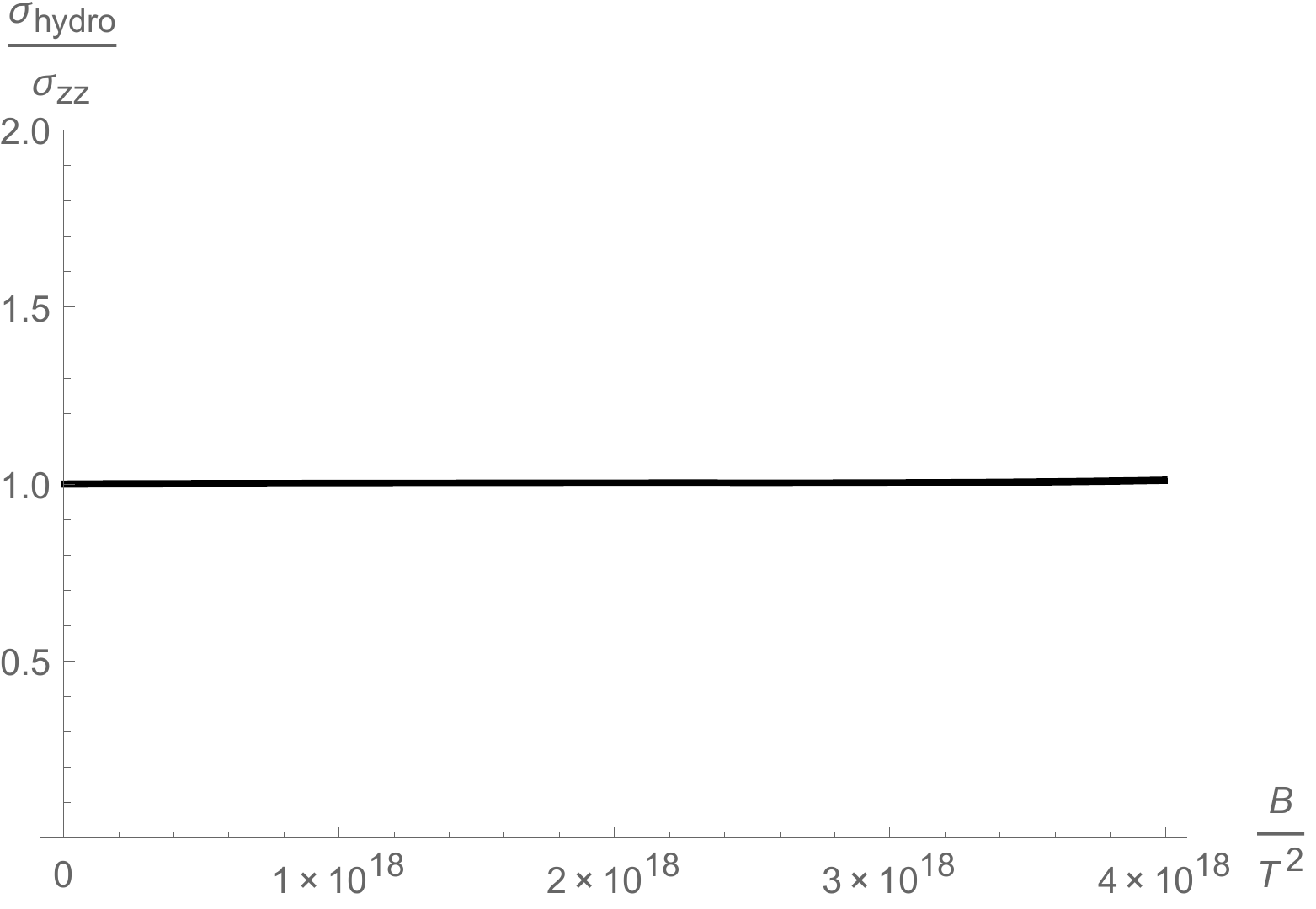} \\
  \end{tabular}
  \caption{\small The ratio of the leading order contribution in the hydrodynamic formula for the longitudinal magnetoconductivity (\ref{hydroformula}) over the leading order contribution in the analytic formula ${32\alpha^2 B^2}/{n_0 \sqrt{h_0} q^2 \phi^2(r_0)}$ for two fixed values of $M/T=0.005\pi\text{ (left)},~0.00005\pi$ (right), $\lambda=200$ and $\alpha=1$.}    \label{agreeah}
\end{figure}

In Fig.~\ref{fig:chi5}, we plot the dependence of $\chi_5$ and its scaling exponent $\gamma_{\chi_5}$ at large $B$ ($\chi_5\simeq c_{\chi_5} B^{\gamma_{\chi_5}}$) on $B/T^2$ for two fixed values of $M/T=0.005\pi\text{ (top)},~0.00005\pi$ (bottom), $\lambda=200$ and $\alpha=1$. Note that in the figure, we defined $\gamma_{\chi_5}=B \chi_5'/\chi_5$, which only has the meaning of the scaling exponent when it reaches a constant in a certain region of $B$. We can see that $\chi_5$ is a monotonically increasing function of $B$ and at $B/T^2\to\infty$ $\chi_5$ grows linearly in $B$, which is the same as the probe limit result. With the scaling behaviors of $\tau_5$ and $\chi_5$ we can see that the hydrodynamic formula also predicts a linear in $B$ behavior for the longitudinal DC magnetoconductivity at $B/T^2\to\infty$. We also checked numerically that the leading order contribution in the hydrodynamic formula (\ref{hydroformula}), i.e. the second term $(8\alpha B)^2\frac{\tau_5}{\chi_5}$ agrees with the leading order contribution in the analytic formula ${32\alpha^2 B^2}/{n_0 \sqrt{h_0} q^2 \phi^2(r_0)}$ as can be seen from Fig.~\ref{agreeah}. This shows that in this backreacted holographic system with axial charge dissipation, the hydrodynamic formula is still valid as long as $\tau_5$ is large enough to stay in the hydrodynamic regime, while $B/T^2$ can be infinitely large, which is outside the hydrodynamic regime.

\section{Conclusion and discussion}
\label{sec5}

In this paper, we considered the backreaction effects of the magnetic field to the holographic longitudinal magnetoconductivity for zero charge and axial charge density chiral anomalous systems. Backreaction effects are important at large $B/T^2$ and large backreaction strength $\lambda$. In the case without axial charge dissipation, the longitudinal magnetoconductivity has a pole in the imaginary part at $\omega=0$. The small frequency result deviates from the probe limit at larger $B/T^2$ region. At $B/T^2\to \infty$, we instead work in the zero temperature limit and find that the imaginary part of the small frequency longitudinal magnetoconductivity coincides with the probe limit result while the real part of the DC longitudinal magnetoconductivity diverges for backreaction strength $\lambda$ larger than a critical value $\lambda_c=(32\alpha)^2$, in contrast to being zero in the probe limit. In the case with axial charge dissipation, the negative magnetoresistivity behavior still exists after including backreactions. At large $B/T^2$ the DC longitudinal magnetoconductivity becomes linear in $B$, which deviates from the exact $B^2$ behavior for the probe limit. Surprisingly we also found that for both cases the hydrodynamic formula for the small frequency longitudinal magnetoconductivity obtained in \cite{Landsteiner:2014vua} still gives the holographic result at zero temperature, which is already out of the hydrodynamic regime.

The calculations in this paper are a first step to the study of holographic negative magnetoresistivity for finite charge and axial charge density systems, where the backreactions of the gauge fields are important to the gravity background. At finite charge density, momentum relaxation is needed in order to have a finite DC longitudinal magnetoconductivity, and at finite axial charge density, energy dissipation will be needed. The next step in this direction would be to add momentum dissipations in the holographic system \cite{{Hartnoll:2012rj},{Horowitz:2012ky},{Liu:2012tr},{Donos:2012js},{Donos:2013eha},{Andrade:2013gsa},{Vegh:2013sk},{Davison:2013txa},{Donos:2014uba}} at finite charge density and compare the holographic result with the hydrodynamic formula. At finite chemical potential and a finite magnetic field background, there exists an instability to spatially modulated phases as shown in \cite{Ammon:2016szz}, which possibly leads to much richer magnetotransport behavior. We will report the study of magnetoresistivity in holographic finite density chiral anomalous systems in the future work.

It is still an open question how to add energy dissipations in holography. At finite axial charge density, it would be interesting to check if there is indeed still a pole at $\omega=0$ after including momentum and axial charge dissipations. Another interesting question is to study the axial charge relaxation and momentum relaxation time from the memory matrix formalism \cite{{Hartnoll:2012rj},Lucas:2015pxa} in the hydrodynamic regime for chiral anomalous systems with a background magnetic field and also check it in strongly coupled holographic systems. Finally, as was found in \cite{Ong}, chiral anomaly also
induces strong suppression of the thermopower in a chiral anomalous system. It would be interesting to study this effect from both the hydrodynamic and holographic point of view.

\subsection*{Acknowledgments}
We would like to thank Rong-Gen Cai, Sean Hartnoll, Karl Landsteiner, Yan Liu, Koenraad Schalm and Jan Zaanen for useful discussions. The work of Y.W.S. was supported by the European Union through a Marie Curie Individual Fellowship MSCA-IF-2014-659135.  The work of Q.Y. was supported by National Natural Science Foundation of China (No.11375247 and No.11435006). This work was also supported in part by the Spanish MINECO's ``Centro de Excelencia Severo Ochoa" Programme under grant SEV-2012-0249. Q.Y. would like to thank the hospitality of IFT during the completion of this work.

\appendix
\section{Zero temperature background solutions with axial charge dissipation}

In this appendix, we present the zero temperature background solutions in the case with axial charge dissipations in the presence of a background magnetic field. We consider the following action
\bea
S&=&\int d^5x \sqrt{-g}\bigg[\frac{1}{2\kappa^2}\Big(R+12\Big)-\frac{1}{4 e^2}\mathcal{F}^2-\frac{1}{4 e^2}F^2+\frac{\alpha}{3}\epsilon^{\mu\nu\rho\sigma\tau}A_\mu \Big(F_{\nu\rho} F_{\sigma\tau}+3 \mathcal{F}_{\nu\rho} \mathcal{F}_{\sigma\tau}\Big)\nonumber\\&&~~~-(D_\mu\Phi)^*(D^\mu\Phi)-m^2\Phi^*\Phi-\frac{\eta}{2}(\Phi^*\Phi)^2\bigg],
\eea
where we have introduced an $\eta |\Phi|^4/2$ term for the convenience of analytic calculation at zero temperature, which does not affect the qualitative properties of transport coefficients.

The equations of motion are
\bea\label{eq:eomtwou1-1}
R_{\mu\nu}-\frac{1}{2}g_{\mu\nu}\Big(R-12 -\frac{\kappa^2}{2 e^2}(\mathcal{F}^2+F^2)-(D_\mu\Phi)^*(D^\mu\Phi)-m^2\Phi^*\Phi-\frac{\eta}{2}(\Phi^*\Phi)^2&\Big)& \nonumber \\
-\frac{\kappa^2}{e^2} \mathcal{F}_{\mu\rho}\mathcal{F}_{\nu}^{~\rho}-\frac{\kappa^2}{e^2} F_{\mu\rho}F_{\nu}^{~\rho}-\kappa^2 ((D_{\mu} \Phi)^* D_{\nu} \Phi+(D_{\nu} \Phi)^*D_{\mu} \Phi)=&0&\\
\nabla_\nu \mathcal{F}^{\nu\mu}+2\alpha\epsilon^{\mu\tau\beta\rho\sigma} F_{\tau\beta}\mathcal{F}_{\rho\sigma}=&0&\,,\\
\label{eq:eomtwou1-2}\nabla_\nu F^{\nu\mu}+\alpha\epsilon^{\mu\tau\beta\rho\sigma} \big(F_{\tau\beta}F_{\rho\sigma}
+\mathcal{F}_{\tau\beta}\mathcal{F}_{\rho\sigma}\big)+i q\big(\Phi (D^\mu\Phi)^*-\Phi^*(D^\mu\Phi)\big)=&0&\,,\\
\label{eq:eomtwou1-3}D_\mu D^\mu\Phi-m^2\Phi-\eta\Phi^{*2}\Phi=&0&\,.
\eea
The assumption for the background solutions is
\be
ds^2=-f(r) dt^2+\frac{d r^2}{f(r)}+n(r) (dx^2+dy^2)+h(r) dz^2,
\ee and \be V_{\mu}=(0,~0,~B y,~0,~0),~~
A_{\mu}=(0,~0,~0,~0,~0),~~\Phi=\phi(r). \ee The equations become
\bea
\frac{f'h'}{2 f h}+\frac{f' n'}{f n}+\frac{h' n'}{h n}+\frac{n'^2}{2 n^2}-\lambda {\phi'}^2-\frac{12}{f}+\frac{\lambda B^2}{2 f n^2}+\frac{\lambda m^2 \phi^2}{f}+\frac{\lambda \eta\phi^4}{2f}=&0&,\\
\frac{f''}{f}-\frac{ n''}{n}+\frac{h'}{2h}\bigg(\frac{f'}{f}-\frac{n'}{n}\bigg)-\frac{\lambda B^2 }{ f n^2}=&0&,\\
\frac{f''}{2f}+\frac{n''}{n}+\frac{n'}{n}\bigg(\frac{f'}{f}-\frac{n'}{4 n}\bigg)-\frac{6}{f}+\frac{\lambda B^2}{4  f n^2}+\frac{\lambda m^2 \phi^2}{2 f}+\frac{\lambda \eta\phi^4}{4f}+\frac{\lambda {\phi'}^2}{2}=&0&\\ \phi''+\phi'\bigg(\frac{f'}{f}+\frac{n'}{n}+\frac{h'}{2 h}\bigg)-\bigg(\frac{m^2\phi}{f}+\frac{\eta \phi^3}{ f}\bigg)=&0&,
\eea
where $\lambda=2\kappa^2/e^2$ and we have rescaled $e \phi \to \phi$ and $\eta/e^2\to \eta$. At zero temperature, an exact solution to the equations above is $AdS_3\times R^2$ with a constant scalar
\bea
ds^2&=&-3(1+\frac{3\lambda}{8\eta})r^2dt^2+\frac{1}{3(1+\frac{3\lambda}{8\eta})} \frac{dr^2}{r^2}+r^2 dz^2+\frac{B\sqrt{\lambda}}{2\sqrt{3(1+\frac{3\lambda}{8 \eta}})}(dx^2+dy^2),\\ \nonumber
\phi&=&\sqrt{\frac{3}{\eta}}.\eea

To flow this solution from the horizon to asymptotic $AdS_5$ we need to find appropriate irrelevant perturbations. Thus up to the first order in perturbations the near horizon solution becomes
\bea
 f&=&3\bigg(1+\frac{3\lambda}{8\eta}\bigg)r^2(1+f_1 r^\beta+\cdots),\\
n&=&\frac{B\sqrt{\lambda}}{2\sqrt{3(1+\frac{3\lambda}{8 \eta})}}\bigg(1-\frac{1}{14}\big(19+2\sqrt{57}\big)f_1 r^\beta\bigg),\\
h&=&r^2(1+f_1 r^\beta+\cdots),\\
\phi&=&\sqrt{\frac{3}{\eta}}(1+f_2 r^\xi+\cdots),
\eea
where $\beta=\frac{1}{3}(\sqrt{57}-3)$ and $\xi=\sqrt{1+\frac{2}{1+\frac{3\lambda}{8\eta}}}-1$. $f_1$ and $f_2$ are two free parameters which can be tuned to get different values of physical $B$ and $M$.

At zero temperature, the equations for the perturbations $v_z$, $\phi_2$ and $a_t$ are the same as equations (\ref{phi2eqns}) of the finite temperature case. The $\phi^4$ term appears in the equation of motion for $\phi_2$ but does not change the equation of motion for $a_r$. When we derive the equation of motion for $\phi_2$ from the three equations (\ref{phi2eqns}), the $\phi^4$ term will arise automatically from the equation of motion of the background scalar field. At zero temperature, in the near horizon region it is difficult to solve for the near horizon behavior of the three fields $v_z$, $a_t$ and $\phi_2$ at $r\ll 1$ while $w$ can be smaller or bigger than $r$. However, we can get the near horizon behavior at $r\ll w$
\bea v_z&\simeq& v_{z0}\sqrt{r} e^{\frac{i \omega}{3(1+\frac{3\lambda}{8\eta}) r}}(1+\cdots),\\ a_t&\simeq& a_{t0}r^{3/2}e^{\frac{i \omega}{3(1+\frac{3\lambda}{8\eta}) r}}(1+\cdots),\\ \phi_2&\simeq& \frac{\phi_{20}}{\sqrt{r}}e^{\frac{i \omega}{3(1+\frac{3\lambda}{8\eta}) r}}(1+\cdots) ,\eea
where \be \phi_{20}=\bigg(-\frac{8\alpha v_{z0} }{q }\sqrt{\frac{3}{\lambda}+\frac{9}{8\eta}}+\frac{i  \omega a_{t0}}{6(1+\frac{3\lambda}{8\eta})  q}\bigg)/\sqrt{\frac{3}{\eta}},\ee and $\cdots$ represent subleading order corrections at order $r^\eta$, $r$, $r^\beta$ and so on.


With these boundary conditions in principle we can solve the zero temperature case numerically and the result would only depend on $B/M^2$, $\alpha$ and $\lambda$. This corresponds to the $B/T^2\to \infty$ and $M/T\to\infty$ limit. As we are more interested in the small $M/T$ while large $B/T^2$ limit, which we already obtained in the finite temperature section, we will not study the zero temperature longitudinal magnetoconductivity here.





\end{document}